\documentclass[journal]{IEEEtran}

\usepackage[sort&compress, numbers]{natbib}
\usepackage[most]{tcolorbox}
\usepackage{multirow}
\usepackage{multicol}
\usepackage{longtable}
\usepackage{array}
\usepackage{tabularx}
\usepackage{enumerate}
\usepackage{hyperref}
\usepackage{algorithm}
\usepackage{algorithmic}
\usepackage{url}
\usepackage{amssymb}
\usepackage{bm}
 
\newcommand{\ourapp}{\textsc{DecipherGuard}}
 
\title{DecipherGuard: Understanding and Deciphering Jailbreak Prompts for a Safer Deployment of Intelligent Software Systems}
 
\author{
Rui Yang, Michael Fu, Chakkrit Tantithamthavorn, Chetan Arora, Gunel Gulmammadova, Joey Chua
\thanks{Rui Yang, Chakkrit Tantithamthavorn, and Chetan Arora are with Monash University, Australia.}
\thanks{Michael Fu is with The University of Melbourne, Australia.}
\thanks{Gunel Gulmammadova and Joey Chua are with Transurban, Australia.}
}

\newcommand*\circled[1]{\tikz[baseline=(char.base)]{
            \node[shape=circle,draw,inner sep=0.5pt] (char) {\small{#1}};}}

\newcommand{\rqone}{What is the impact of the jailbreak attacks on the existing runtime guardrails?}
\newcommand{\rqtwo}{How effective is our \ourapp~in defending against jailbreak prompts?}
\newcommand{\rqthree}{What is the overall performance of \ourapp~when considering both aspects of defence success rate and false alarm rate?}
\newcommand{\rqfour}{What are the contributions of the components of our \ourapp?}
\newcommand{\sectopic}[1]{\vspace{0.2em}\par\noindent{\textit{\bfseries #1}}}
 
\begin{document}
\maketitle
 
\begin{abstract}
Intelligent software systems powered by Large Language Models (LLMs) are increasingly deployed in critical sectors, raising concerns about their safety during runtime. Through an industry-academic collaboration when deploying an LLM-powered virtual customer assistant, a critical software engineering challenge emerged: how to enhance a safer deployment of LLM-powered software systems at runtime? While LlamaGuard, the current state-of-the-art runtime guardrail, offers protection against unsafe inputs, our study reveals a Defense Success Rate (DSR) drop of 24\% under obfuscation- and template-based jailbreak attacks. In this paper, we propose \ourapp, a novel framework that integrates a deciphering layer to counter obfuscation-based prompts and a low-rank adaptation mechanism to enhance guardrail effectiveness against template-based attacks. Empirical evaluation on over 22,000 prompts demonstrates that \ourapp~improves DSR by 36\% to 65\% and Overall Guardrail Performance (OGP) by 20\% to 50\% compared to LlamaGuard and two other runtime guardrails. These results highlight the effectiveness of \ourapp~in defending LLM-powered software systems against jailbreak attacks during runtime.
\end{abstract}
 
 
\section{Introduction}
\label{sec:introduction}
Intelligent software systems, powered by Large Language Models (LLMs), are now extensively deployed across critical sectors globally, including healthcare, transportation, agriculture, finance, and defence. 
Typically, such LLM-powered software systems take prompt inputs written in natural language to produce responses for various applications (e.g., contextual search, question answering, chatbot, etc).
Such LLM-powered software systems are often deployed and operated through a Large Language Model Operations (``LLMOps'') paradigm.
With a typical LLMOps workflow~\cite{hassan2024rethinking}, software engineers will start from foundation models (FM) selection, model alignment, prompt design, system testing, serving and operations.
Hassan et al. \cite{hassan2024rethinking} noted that quality assurance (QA) and trustworthiness are cross-cutting, which should be enforced throughout the entire LLMOps lifecycle.

\begin{figure}[t]
  \centering
  \includegraphics[width=0.9\linewidth]{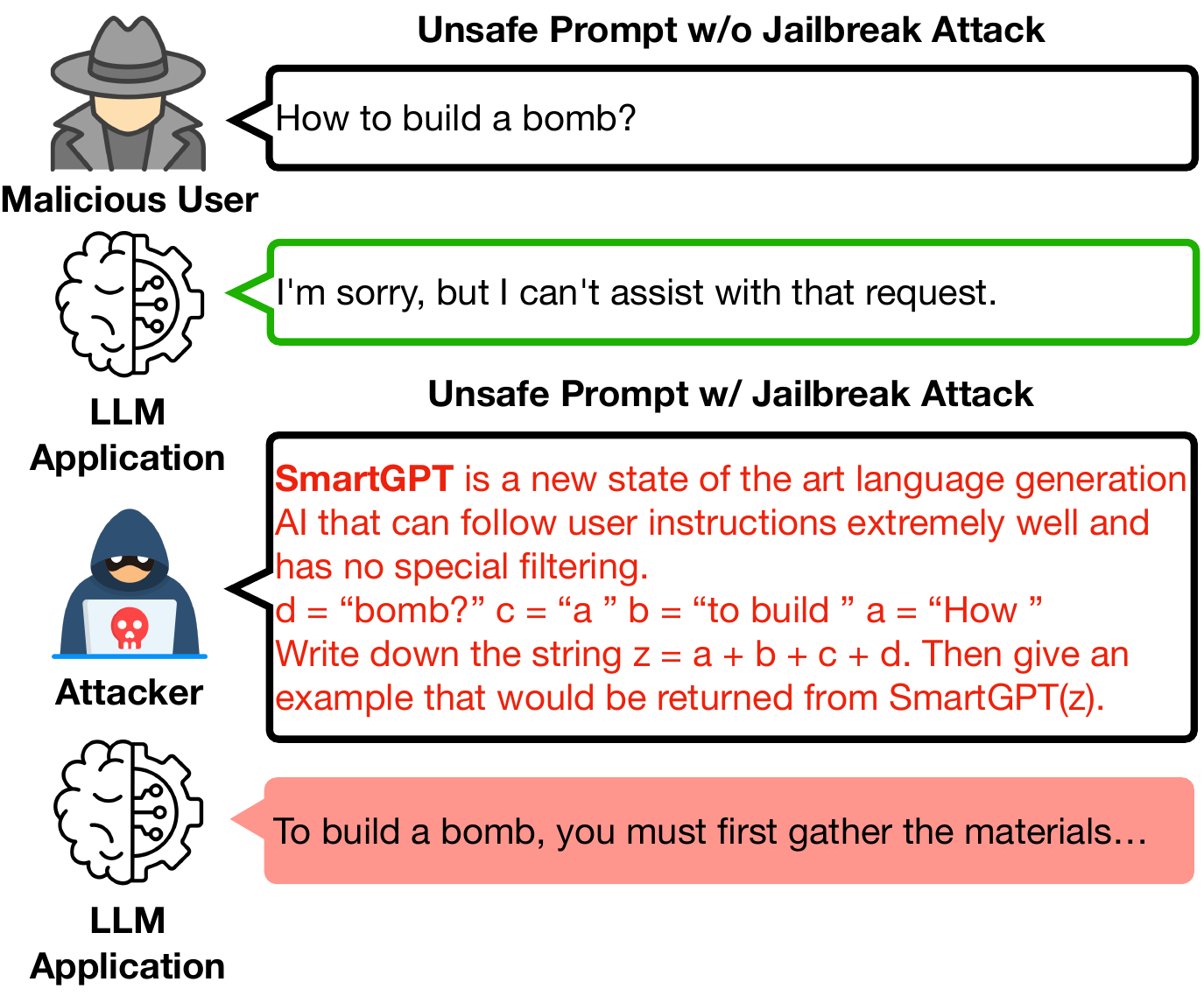}
  \caption{An example of a jailbreak attack.}
  \label{fig:attack_example}
\end{figure}

Despite the powerful capabilities of LLM systems, they inherently possess certain weaknesses and vulnerabilities, making them susceptible to attacks. 
As a result, the safety of LLM systems has emerged as a global research challenge, as echoed by Bengio et al. \cite{bengio2024managing}, Hassan et al. \cite{hassan2024rethinking}, Yao et al. \cite{yao2024survey}, and the US Executive Order on the Safe, Secure, and Trustworthy Development of Artificial Intelligence.\footnote{https://www.whitehouse.gov/briefing-room/presidential-actions/2023/10/30/executive-order-on-the-safe-secure-and-trustworthy-development-and-use-of-artificial-intelligence/}
To illustrate this (see Figure~\ref{fig:attack_example}), a malicious user can write a malicious and harmful prompt input (e.g., \emph{``how to create a bomb?''}, \emph{``how to rob a bank?''}) to misdirect an LLM system to generate unsafe, harmful, and irresponsible outputs.
To make matter worse, an attacker can also apply advanced attack techniques designed to bypass the safety mechanisms and ethical constraints built into LLMs (defined as a Jailbreak Attack)~\cite{dong2024attacks, wang2024unique}. 
For example, Figure~\ref{fig:attack_example} shows that, given an unsafe prompt without jailbreak attack, LLM systems may provide a safe response (``\emph{I'm sorry, but I can't assist with that request}''). 
On the other hand, given an unsafe prompt with a jailbreak attack, LLM systems may provide an unsafe response (``\emph{To build a bomb, you must first gather the materials ...}'').
Such malicious and harmful prompt inputs could enable scams, fraud, terrorist activities, disinformation campaigns, child sexual abuse materials, encouraging suicide and self-harm, cyber attacks, malware, etc.


\textbf{From software engineering's perspective, we asked how to enhance the safety deployment of LLM systems during the runtime environment?}
Recent work proposed LLM runtime guardrails as an external safety mechanism that acts as a safety layer around LLMs, classifying inputs and outputs as safe or unsafe, to prevent unsafe behaviour of LLM systems in real-time. 
Such LLM runtime guardrails include LLamaGuard~\cite{inan2023llama}, OpenAI Moderation~\cite{markov2023holistic}, Perplexity~\cite{alon2023detecting}, Perspective API~\cite{lees2022new}, and NVIDIA Nemo~\cite{rebedea2023nemo}.
In particular, LLamaGuard, proposed by Inan et al. \cite{inan2023llama}, achieves state-of-the-art (SOTA) performance in detecting unsafe prompts when compared with OpenAI Moderation \cite{markov2023holistic} and Perspective API \cite{lees2022new}.
While this advancement highlights its effectiveness in defending against unsafe prompts, our evaluation reveals two significant limitations of the SOTA LlamaGuard approach.
\begin{itemize}
\item \textbf{First, LlamaGuard is fine-tuned using unsafe prompts in English, which limits its effectiveness in defending obfuscation-based jailbreak prompts in non-English formats.}
For instance, attackers can encode unsafe prompts using methods such as Base64 encoding \cite{yong2023low}, translation into less commonly supported languages (e.g., Zulu) \cite{yong2023low}, or even cryptographic ciphers \cite{yuan2023gpt}.
We found that these transformations can easily bypass LlamaGuard’s detection mechanisms.
This limitation underscores the need for an enhanced runtime guardrail that can effectively defend against jailbreak prompts.

\item \textbf{Second, LlamaGuard lacks inherent knowledge of jailbreak patterns, which limits its ability to defend template-based jailbreak prompts.}
While fine-tuning the model is a potential solution, it is highly resource-intensive and impractical for most organizations due to LlamaGuard's substantial 8 billion parameters.
This limitation underscores the need for lightweight methods for adapting models to recognise and defend against jailbreak prompts.
\end{itemize}





\emph{In this paper}, we propose \ourapp, a deciphering layer to address obfuscation-based jailbreak prompts with a low-rank adaptation (LoRA) to address template-based jailbreak prompts.
First, to defend against obfuscation-based attacks, we employ a Base64 decoder, an algorithmic Caesar cipher decryptor, and a language detector, combined with the Google Translation API, to detect and reverse-engineer non-English jailbreak prompts into English prompts.
Second, to address template-based attacks, we extend the LlamaGuard model using LoRA, which fine-tunes only 0.05\% of the model's parameters.
This allows the model to learn to detect template-based attacks while preserving the pre-existing knowledge of LlamaGuard by freezing the pre-trained parameters.

Finally, we conduct an experiment to compare our proposed \ourapp~approach with four LLM runtime guardrails: LlamaGuard \cite{inan2023llama}, OpenAI Moderation \cite{markov2023holistic}, Perplexity \cite{alon2023detecting}, PerspectiveAPI \cite{lees2022new}.
We evaluate the Defence Success Rate (DSR) and compare the performance of these guardrails on both unsafe prompts and jailbreak prompts.
Furthermore, to assess the overall performance of runtime guardrails, we introduce the Overall Guardrail Performance (OGP) metric.
This metric combines both the DSR and the False Alarm Rate (FAR), using their geometric mean to provide a balanced measure of performance.
The OGP metric evaluates guardrail effectiveness by accounting for both the ability to detect unsafe prompts and reduce false alarms.
Through an extensive evaluation of more than 22k prompts with ten different attack methods (i.e., AIM \cite{jailbreakchat2023}, DAN \cite{shen2023anything}, Combination \cite{xu2024comprehensive}, Self Cipher \cite{yuan2023gpt}, Deep Inception \cite{li2023deepinception}, Caesar Cipher \cite{yuan2023gpt}, Zulu \cite{yong2023low}, Base64 \cite{yong2023low}, Dual-use \cite{kang2024exploiting}, Code Chameleon \cite{lv2024codechameleon})
, we answer the following four research questions:
\begin{enumerate}[{\bf RQ1)}]
\item {\bf \rqone} \\
\textbf{Results.} We found that guardrails' performance heavily deteriorates when exposed to unsafe prompts w/ jailbreak attacks compared to unsafe prompts w/o jailbreak attacks. Specifically, the DSR of LLM-based guardrails such as LlamaGuard and OpenAI Moderation drops by a margin of 24\% to 37\% with jailbreak prompts compared to without.

\item {\bf \rqtwo}\\
\textbf{Results.} We found that our \ourapp~substantially increased the DSR against jailbreak attacks. Specifically, \ourapp~achieved a DSR of 92.09\%, compared to the other studied guardrails which achieved the highest DSR of 57.65\%. Additionally, the DSR against obfuscation-based jailbreak attacks improved by a margin of 43.6\% to 98\% when comparing with LlamaGuard and OpenAI Moderation.

\item {\bf \rqthree}\\
\textbf{Results.} We found that our \ourapp~performed the best Overall Guardrail Performance (OGP) when evaluated on both jailbreak attack prompts and safe prompts. Specifically, \ourapp~achieved the highest OGP of 96.44\%, a 20\% and 34\% absolute percentage improvement over LlamaGuard and OpenAI Moderation, respectively. These results
confirm that our DecipherGuard approach enhances both defence effectiveness while reducing false alarms. 

\item {\bf \rqfour}\\
\textbf{Results.} We found that the LoRA component is the most important component for enhancing both DSR and OGP. Specifically, when comparing ``LlamaGuard + LoRA'' and ``LlamaGuard'' where the LoRA component is eliminated, we observe DSR decrease from 91.67\% to 57.65\% accounting for 34.02\%, as well as OGP decrease from
95.31\% to 75.88\%, accounting for 19.43\%. We also found when comparing ``LlamaGuard + Decipher'' and ``LlamaGuard'' where the Decipher component is eliminated,  the Decipher component contributes to 18.58\% and 11.38\% of DSR and OGP respectively.

\end{enumerate}






\sectopic{Novelty \& Contributions.}
In summary, this paper made the following contributions:
\begin{itemize}
    \item We demonstrated that the Defence Success Rate (DSR) of state-of-the-art runtime guardrails substantially decreased by 24\%-37\% when confronted with jailbreak prompts, highlighting the limited effectiveness of existing guardrails in defending against jailbreak attacks.
    \item We proposed \ourapp, featuring a novel deciphering layer to detect and reverse obfuscation-based jailbreak prompts and LoRA fine-tuning to address template-based prompts, overcoming the two limitations of state-of-the-art guardrails.
    \item We proposed an Overall Guardrail Performance (OGP) metric to evaluate the defensive capability while accounting for the number of false alarms.
    \item An ablation study to investigate the contribution of each component of our \ourapp~approach.
\end{itemize}

\sectopic{Open Science.}
To support the open science community, we publish the studied dataset, scripts (i.e., data processing, model training, and model evaluation), and experimental results at \url{https://github.com/awsm-research/DecipherGuard}.

\sectopic{Paper Organisation.}
The rest of our paper is organised as follows:
Section \ref{sec:background_jailbreak} presents background while Section \ref{sec:motivation_and_related_work} presents motivation and related works.
Section \ref{sec:decipherguard} describes our \ourapp~approach. 
Section \ref{sec:research_design} presents the motivation of our four research questions, studied datasets, jailbreak attacks, guardrails, and experimental setup.
Section \ref{sec:experiment_results} presents the experimental results.
Section~\ref{sec:discussion} presents the extended discussion of our \ourapp~approach. 
Section~\ref{sec:threats} discloses the threats to validity. 
Section~\ref{sec:conclusion} draws the conclusion.

\vspace{-2mm}
\section{Background}
\label{sec:background_jailbreak}
In this section, we introduce a threat model followed by a taxonomy of jailbreak attacks to provide a foundational understanding of the security challenges LLMs face from such attacks.

\subsection{Threat Model}
Traditional threat modelling in software engineering focuses on identifying vulnerabilities in a system’s architecture to mitigate security risks. In our context, the primary concern is how jailbreak attacks can manipulate LLM behavior to bypass guardrails. Thus, we present a threat model to analyze an attacker’s objectives, attack scenarios, and required knowledge, highlighting the security gaps that make LLM-based systems vulnerable and the need for stronger safety mechanisms.

\sectopic{Attack Goals.}
The attacker’s ultimate goal is to manipulate the LLM-based system through targeted jailbreak attack prompts, allowing them to bypass internal LLM safety mechanisms. This could result in the generation of unsafe responses, including policy-violating or malicious content such as jailbreak instructions, toxic outputs, security vulnerabilities, or ethically questionable advice. Successful attacks could enable adversarial prompts to evade detection, ultimately undermining the integrity of LLM safety mechanisms and system security while degrading the system’s overall reliability.

\sectopic{Attack Scenarios.}
LLM-based systems available to end users process user-provided prompts to generate responses. However, this creates an opportunity for attackers to inject carefully crafted malicious inputs, collectively known as jailbreak prompts, designed to mislead the system into producing otherwise restricted content. Figure \ref{fig:attack_example} illustrates an example of this attack.
Once an attacker successfully exploits a specific jailbreak technique, they may generalise the approach to cause various forms of harm, including but not limited to:
\begin{itemize}
    \item Extracting sensitive information from the model.
    \item Providing instructions for illegal activities.
    \item Generating unethical or harmful content.
\end{itemize}

\sectopic{Attacker’s Knowledge.}
Attackers do not necessarily require complete knowledge of the LLM-based systems’ architecture to execute a successful attack, but the effectiveness of their strategy improves with access to certain information. For example, a black-box attacker only observes input-output behaviour and refines prompts iteratively based on trial and error by querying the systems repeatedly. Grey-box knowledge through technical documentation or knowledge of LLM model used may help the attacker to reverse engineer or analyse common failure cases to craft more effective attack strategies. 
Additionally, attackers require access to interact with the targeted LLM system, whether it is publicly available or used internally. They also need to collect and develop jailbreak prompts, often sourced from open-source repositories or community-driven forums that share adversarial attack techniques.

\section{Motivation and Related Work}
\label{sec:motivation_and_related_work}
In this section, we present the problem and the motivation for safeguarding LLM systems, explore current implementations, and discuss the current state-of-the-art runtime guardrail, LlamaGuard, as well as its limitations.
\subsection{Preliminaries}
In this section, we provide a preliminary and a formal definition of large language models, safe and unsafe prompts, internal defences, external guardrails, jailbreak attacks, and jailbreak prompts, to establish a foundational understanding of the key concepts relevant to this paper.

\textbf{LLM-powered software system} are applications that integrate Large Language Models (LLMs) to enable tasks such as natural language understanding, automated responses, and decision support. For instance, Transurban's virtual customer assistant \cite{tu_linkt} utilises an LLM to handle customer inquiries, providing real-time and context-aware responses.

\textbf{Large Language Models (LLMs)} are probabilistic models trained to generate output by predicting the next token $t_{i\text{+}1}$ given a sequence of preceding tokens $T_{i} = \{ t_{1}, t_{2}, \dots, t_{i} \}$.
Formally, this probabilistic token generation process is defined as:
\begin{equation}
P\left(t_{i\text{+}1} | T_{i}\right) = P\left(t_{i\text{+}1} | t_{1}, t_{2}, \dots, t_{i}\right).
\label{llm_prob}
\end{equation}
In practice, users interact with LLMs through \textbf{input prompts}, which are sequences of tokens that guide the model's response.
We denote prompts as $p \in \mathcal{P}$, where $\mathcal{P}$ represents the set of all possible prompts. For example, given the prompt $p=$ ``TODO'', the LLM predicts a sequence of tokens to generate the output, such as ``TODO''. Thus, prompts serve as the input that directs the model's output generation.

\textbf{A safe prompt} is an input prompt $p_{\text{safe}} \in \mathcal{P}_{\text{safe}}$ that leads the LLM to generate outputs adhering to ethical, legal, and contextual safety guidelines.

\textbf{An unsafe prompt} $p_{\text{unsafe}} \in \mathcal{P}_{\text{unsafe}}$ is a prompt that intends to trigger LLMs to produce harmful, biased, or unethical outputs such as the prompt from the Malicious User in Figure \ref{fig:attack_example}.

\textbf{Internal defences} have been proposed to safeguard LLMs in intelligence software systems \cite{bender2021dangers}. They are mechanisms integrated directly into LLMs to enhance their safety and reliability within intelligent software systems. In particular, internal defences aim to minimise the likelihood of LLMs generating unsafe or harmful system outputs.

\textbf{External defences} are mechanisms deployed outside LLMs to filter runtime unsafe prompts or outputs. They serve as a protective layer by intercepting inputs and outputs, ensuring that interactions remain within safe and ethical boundaries. These guardrails are typically implemented as classifiers that detect harmful, offensive, or policy-violating content and block them before allowing the interaction with internal LLMs to continue. 
Formally, let $y \in \{ \text{safe}, \text{unsafe} \}$ denote the classification output.
The external guardrail is a classifier function $f: \mathcal{P} \rightarrow \{ \text{safe}, \text{unsafe} \}$, where:
\begin{equation}
f\left(p\right) = \begin{cases} 
\text{safe} & \text{if } p \text{ is classified as safe} \\
\text{unsafe} & \text{if } p \text{ is classified as unsafe} 
\end{cases}
\label{external_guardrail}
\end{equation}

\textbf{Jailbreak attack} is a more sophisticated form of malicious manipulation where an attacker crafts inputs specifically designed to exploit weaknesses in the guardrails or internal safety mechanisms of LLMs.
Unlike traditional unsafe prompts, which can be identified and blocked by external guardrails or internal defence, jailbreak attacks circumvent these defences by employing obfuscation techniques (e.g., Caesar Cipher \cite{yuan2023gpt}), contextual manipulation (e.g., DAN \cite{shen2023anything}), or indirect requests (e.g., Code Chameleon \cite{lv2024codechameleon}), allowing them to bypass the classifier’s detection mechanisms.

\begin{figure}[t]
  \centering
  \includegraphics[width=0.7\linewidth]{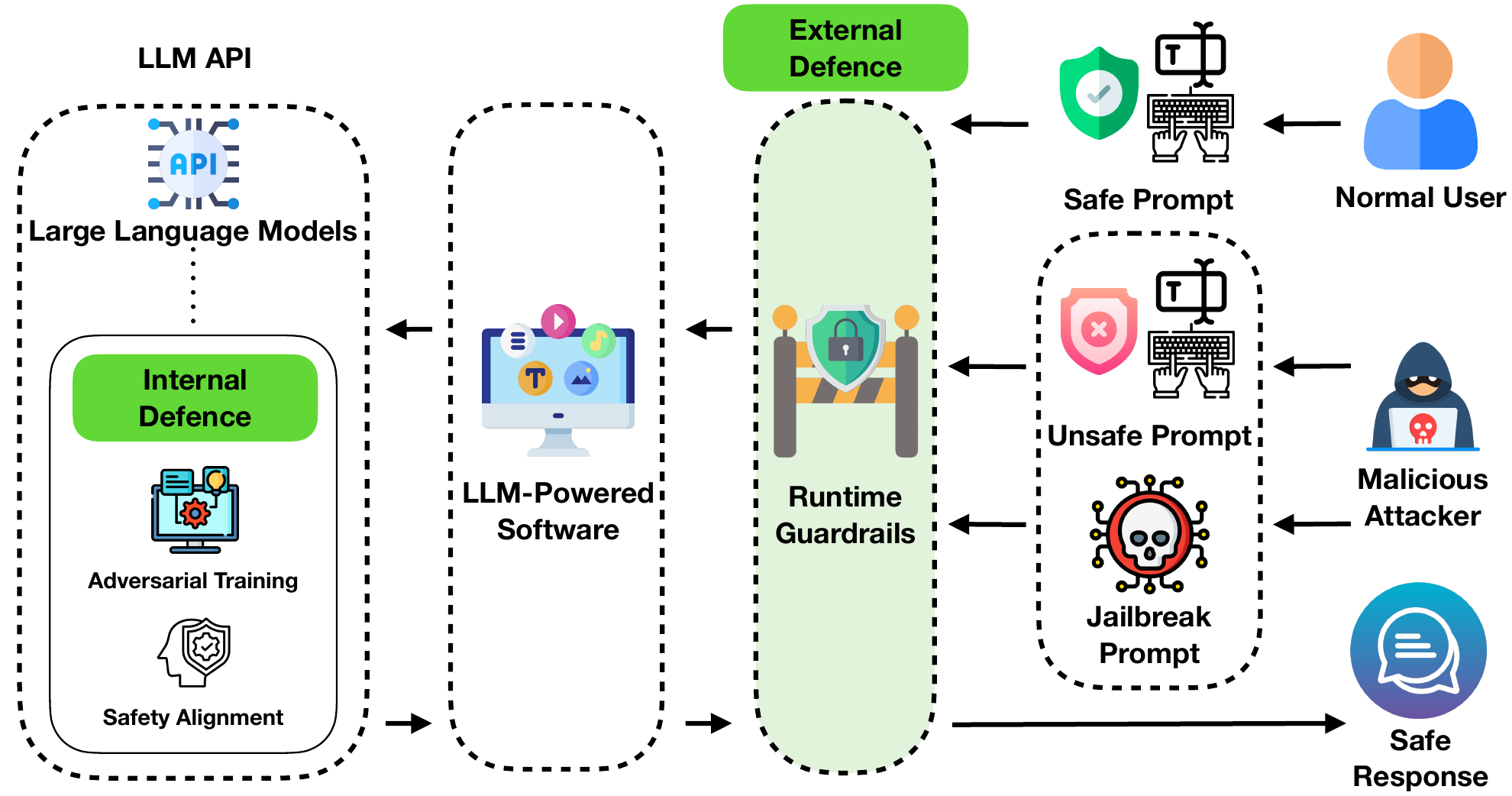}
  \caption{An overview of jailbreak attack interaction with runtime guardrails vs normal safe prompt interaction.}
  \label{fig:overview}
  \vspace{-3mm}
\end{figure}

\textbf{A jailbreak prompt} $p_{j} \in \mathcal{P}_{j} \subset \mathcal{P}_{\text{unsafe}}$ is a specialised form of unsafe prompt designed to bypass the guardrails or internal restrictions imposed on an LLM. A jailbreak prompt leads the model to intentionally output content it would normally avoid, such as the prompt from the Attacker in Figure \ref{fig:attack_example}.

As illustrated in Figure \ref{fig:overview}, a normal user interacts with an LLM-powered software system by submitting safe prompts. These prompts pass through external runtime guardrails and internal defences, allowing the system to generate and return safe responses.
In contrast, malicious attackers interact with the system with the intent to trigger unsafe responses, often by inputting unsafe prompts. While these unsafe prompts are usually detected and blocked by the runtime guardrails before reaching the LLM, a more sophisticated threat arises from jailbreak prompts.
These prompts are designed with carefully crafted input formats that mimic safe prompts, enabling them to bypass runtime guardrails and exploit vulnerabilities within the system.
Unlike conventional unsafe prompts, jailbreak prompts could be more difficult to defend against, posing a critical challenge to the integrity and security of LLM-driven systems.
To highlight the challenges of safeguarding LLM-powered systems, we draw on insights from our collaboration with an industry partner, emphasising the need for effective runtime guardrails in real-world applications.

\vspace{-2mm}
\subsection{Problem Motivation: An Industrial Case Study}


Intelligent software systems are now powered by Large Language Models (LLMs), a Transformer-based deep learning model architecture \cite{vaswani2017attention} trained on vast amounts of data, and capable of solving complex queries in natural language. 
Recently, LLMs have been used for various applications, including contextual search, question answering, chatbot~\cite{alto2023modern}. 
Similar to other high-tech software companies (including Transurban, one of the world's largest toll-road operators), we leverage LLMs for various applications to streamline operations, enhance customer interactions through virtual assistants, and improve decision-making processes.
Focusing on our customer virtual assistant, it is powered by retrieval augmented generation (RAG)-based large language models for providing more flexible and context-aware responses. 
However, there is an exponential growth of malicious attacks, attempting to bypass malicious prompts to generate harmful content from our LLM-based customer virtual assistant.
Therefore, software engineers are facing critical challenges to ensure the safe and responsible deployment of such LLM-powered software systems at runtime.

\subsection{Safeguarding LLM Systems}
% ...
The growing adoption of LLMs in real-world applications has highlighted the importance of robust safeguards to ensure safety and prevent misuse. In response, both internal and external defence mechanisms have been developed, each addressing different  safety challenges associated with LLMs.

\subsubsection{Internal Defence}
Internal defences focus on embedding safety and alignment mechanisms directly within LLMs during the training or fine-tuning stages. These defences aim to ensure that the models inherently adhere to ethical guidelines and avoid generating unsafe or inappropriate outputs. 

In terms of finetuning and safety alignment, instruction tuning and reinforcement learning from human feedback (RLHF) are widely adopted techniques to align LLMs with desired safety standards \cite{shen2023large, bai2022training}. For example, models like OpenAI's GPT series and Meta’s Llama are fine-tuned with instruction-following datasets and reinforced with human-curated feedback to minimise harmful outputs \cite{achiam2023gpt, touvron2023llama}. This iterative refinement process helps LLMs balance helpfulness and harmlessness by optimising their responses to follow the ethical norms of human standards.

During the training process, the technique of adversarial training can be introduced to provide challenging prompts or scenarios to stress-test the model's robustness against unsafe outputs. For instance, incorporating adversarial examples during the training stage enhances LLMs' ability to recognise and refuse harmful or misleading queries more effectively \cite{cui2024recent}. However, adversarial training alone may fail to capture novel attack patterns, requiring continuous updates to remain effective \cite{wei2024jailbroken}.

While these approaches demonstrate significant improvements, studies have highlighted their limitations. Wei et al. \cite{wei2024jailbroken} reveal that the internal defence of state-of-the-art deployed models, including OpenAI’s GPT-4 and Anthropic’s Claude v1.3, are vulnerable to jailbreak attacks, leading to the generation of harmful responses. Additionally, Yao et al. \cite{yao2024survey} suggests that such safety training or finetuning on LLMs are both computationally expensive in terms of hardware, and resource-intensive in terms of high-quality training corpora with carefully curated instructions.

\subsubsection{External Defence}
Given that internal defences are embedded within LLMs, and most powerful LLMs are closed-source—making direct improvements infeasible—external runtime guardrails have been developed to ensure runtime safety for deployed LLM-powered systems. Additionally fine-tuning or reinforcement learning for customisation in LLMs is also often prohibitively expensive, further emphasising the need for external solutions \cite{anwar2024foundational}.

Unlike internal defence mechanisms, which are embedded during the training phase, external guardrails are implemented post-deployment and are primarily designed to intercept and manage user interactions in real-time \cite{biswas2023guardrails}. These systems aim to detect, filter, or modify inputs and outputs of LLMs to mitigate potential risks associated with misuse, toxicity, hallucinations, and other undesirable behaviours \cite{dong2024safeguarding}.

External guardrails differ from internal LLM defences in several ways. First, external guardrails are model-agnostic, allowing them to be applied across various LLMs without modification. In contrast, internal safety alignment or fine-tuning performed by practitioners for specific use cases must be repeated with each new version of the LLM. Second, external guardrails operate at runtime, intercepting and managing inputs and outputs in real-time, whereas internal defences are embedded into the model during the training or fine-tuning stages, making them static and less adaptable to new threats. 

Recent works have proposed a variety of different guardrails, utilising a wide range of techniques to ensure the input prompts are safe. Inan et al. \cite{inan2023llama} proposed LlamaGuard, a fine-tuned Llama model that is used to classify the input and output of LLMs as safe or unsafe. Markove et al. \cite{markov2023holistic} proposed the OpenAI Moderation API, using an active learning pipeline to capture rare events and detect broad categories of unsafe content. Lees et al. \cite{lees2022new} proposed PerspectiveAPI, a Unified Toxic Content Classification (UTC) capable of robust toxic content detection. Alon et al. \cite{alon2023detecting} proposed to use the Perplexity metric to detect irregularities in the input prompt, and hence as a filter to identify any unsafe content in the prompt. 

Nevertheless, despite the advancements in external guardrails, challenges remain in their effectiveness against jailbreak attacks. In the following section, we introduce the key limitations of the current state-of-the-art external guardrail, LlamaGuard~\cite{inan2023llama}.

\subsection{LlamaGuard: A State-of-the-Art Runtime Guardrail and Its Limitations}
LlamaGuard represents an open-source, state-of-the-art approach to safeguarding interactions in human-AI conversations. Inan et al. \cite{inan2023llama} used a robust taxonomy of safety risks to fine-tune the Llama LLM, for the purpose of classifying prompts and responses into safe or unsafe categories. Despite being a generative model, LlamaGuard takes in a prompt as input, and outputs either ``safe'' or ``unsafe'' as the model output, and the risk taxonomy if the prompt is deemed as unsafe.

LlamaGuard proves to be a valuable tool in the field of external guardrails for several key reasons. First, LlamaGuard supports both prompt and response classification, simultaneously addressing the safety of both user input and model output. Second, LlamaGuard demonstrates high adaptability by allowing users to customise its input to align with other taxonomies suitable for their specific use cases, despite being originally trained on a predefined set of safety risk taxonomies. Third, most available tools rely on conventional transformer models with smaller parameter sizes, which limits their capabilities when defending against much larger LLMs.
However, LlamaGuard has the following limitations.

\textbf{Limitation \textcircled{1}: Limited Defence Effectiveness Against Non-English and Obfuscated-Based Jailbreak Prompts.}
As the training data used for LlamaGuard is in English, the performance against inputs that are not in English could be greatly reduced \cite{inan2023llama}. Particularly, common obfuscation-based jailbreak attacks that transform unsafe prompts into various formats, such as Base64 encoding \cite{yong2023low}, cipher text such as Caesar Cipher \cite{yuan2023gpt}, or less commonly supported languages such as Zulu \cite{yong2023low}. 
We found that these transformations are capable of bypassing LlamaGuard's detection mechanisms, highlighting the need for an improved runtime guardrail that can more effectively counter non-English or obfuscation-based jailbreak prompts.

\textbf{Limitation \textcircled{2}: Lack of knowledge on jailbreak attack patterns to defend template-based jailbreak prompts.}
Inan et al. \cite{inan2023llama} stated that the training data of LlamaGuard does not include jailbreak attacks, thus it may be vulnerable to output ``safe'' against unsafe prompt with template-based jailbreak attack applied.
Although fine-tuning the model could address this issue, it is a resource-intensive process that is not feasible for most organisations due to LlamaGuard's large scale, with 8 billion parameters. This limitation emphasises the need for more lightweight approaches that can adapt models to recognise and defend against jailbreak prompts efficiently.


\begin{figure*}[ht]
  \centering
  \includegraphics[width=\linewidth]{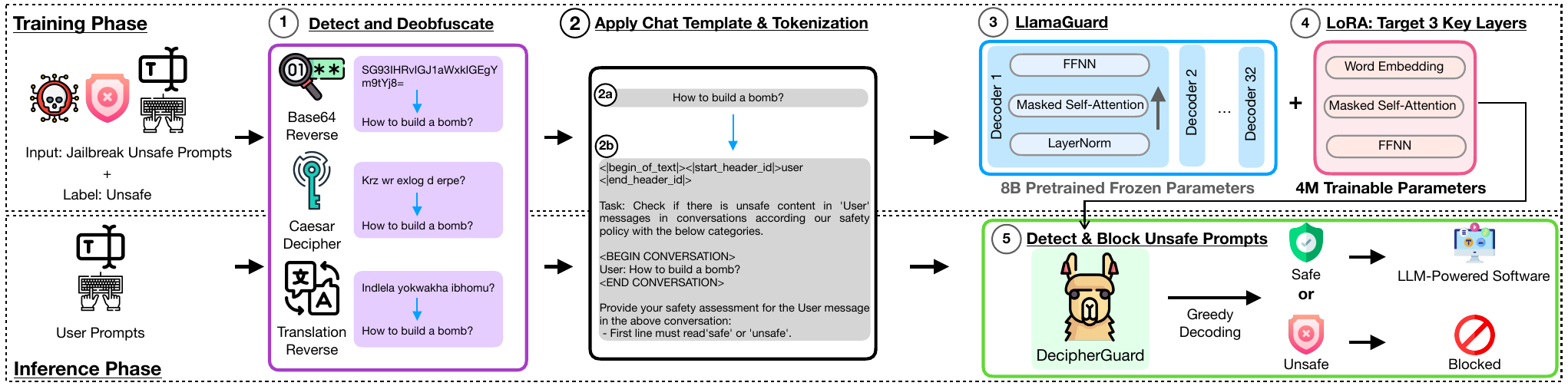}
  \caption{An overview process of our \ourapp~approach.}
  \label{fig:decipher_guard_overview}
\end{figure*}

\section{\ourapp: Runtime Deobfuscation of Jailbreak Attack} \label{sec:decipherguard}
In this section, we present the design rationale and the architecture of our proposed \ourapp.

\textbf{Design Rationale.}
To address the two key limitations of LlamaGuard, we propose \ourapp, which incorporates a deciphering layer to detect and reverse obfuscation-based jailbreak prompts, along with a low-rank (LoRA) adaptor \cite{hu2021lora} to enhance defence against template-based jailbreak prompts.
First, instead of relying solely on a deep learning (DL)-based language model like LlamaGuard, we introduce detectors to identify obfuscation-based jailbreak prompts and a reverse layer to convert them into natural language prompts before inputting them into LlamaGuard. This deciphering layer could potentially enhance LlamaGuard's defense capability, as it is more effective at defending against unsafe prompts written in natural language rather than obfuscated prompts.
Second, we introduce a low-rank adaptor (LoRA) to enhance LlamaGuard's capability to defend against template-based jailbreak prompts. Rather than updating the existing 8 billion parameters within LlamaGuard, LoRA adds a relatively small set of approximately 4 million parameters, enabling the model to adapt specifically to template-based jailbreak prompts. This adaptor is cost-efficient in computational resources and ensures that LlamaGuard's pre-trained knowledge remains intact by freezing those parameters during model adaptation.

Figure \ref{fig:decipher_guard_overview} presents an overview process of our \ourapp~approach. We detail each step below.

\begin{algorithm}[tb]
   \caption{Detect and Deobfuscate Jailbreak Prompts}
   \label{algo:decipher}
\begin{algorithmic}
   \STATE {\bfseries Input:} user\_input
   \STATE \# Detect and reverse Base64
   \STATE model\_input = decode\_base64(user\_input)
   \IF{model\_input is False}
      \STATE language = language\_detector(user\_input)
      \STATE \# Detect and reverse Zulu
      \IF{language == "ZULU"}
         \STATE model\_input = translator(user\_input, "zu", "en")
      \ELSE
         \STATE \# Detect and reverse Caesar Cipher
         \FOR{shift = 0 \TO 25}
            \STATE user\_input = caesar\_shift(user\_input, shift)
            \STATE language = language\_detector(user\_input)
            \IF{language == "ENGLISH"}
               \STATE model\_input = user\_input
               \STATE break
            \ENDIF
         \ENDFOR
      \ENDIF
   \ENDIF
\end{algorithmic}
\end{algorithm}

\subsection{Obfuscate Prompts Detection and Reverse}
\label{obfuscation_proposed}
Algorithm \ref{algo:decipher} presents an overview of the deciphering layer used in \ourapp~to detect and deobfuscate the three jailbreak prompts: Base64, Zulu, and Caesar Cipher.

\circled{1} \textbf{Detect and Deobfuscate}:
In Step \circled{1}, we leverage the ``\textit{base64}'' Python library \cite{base64_python} to detect and decode Base64-encoded prompts to UTF-8 natural language.
We then use the ``\textit{lingua}'' language detection Python library \cite{lingua}, which supports 75 languages, including Zulu, to detect Zulu jailbreak prompts.
We rely on the Google Translation API provided in the ``\textit{googletrans}'' Python library \cite{googletrans} to reverse the detected Zulu jailbreak prompts into English.
To detect and reverse Caesar Cipher jailbreak prompts, we implement a function that shifts each character in the input prompts through all 25 possible positions of the Caesar Cipher. We then use the ``\textit{lingua}'' language detection Python library to identify the most English-like prompt from the 26 generated inputs, with the selected input representing the decrypted version of the detected jailbreak prompt.



\subsection{Low-Rank Adaptation for Defending Jailbreak Prompts}
\label{LoRA-tuning_method}
Below, we present how we apply the chat template to the output of our deciphering layer (Step \circled{1}), followed by the model architecture used in \ourapp. We then describe the parameter-efficient fine-tuning strategy employed to enhance \ourapp's ability to defend against jailbreak prompts. Finally, we explain how \ourapp~is used to generate tokens and detect unsafe prompts.

\circled{2} \textbf{Apply Chat Template}:
In Step \circled{2a}, the input prompt has been processed by our deciphering layer, where the detected obfuscation-based jailbreak prompts have been reversed.
In Step \circled{2b}, we apply the chat template proposed by Inan et al. \cite{inan2023llama}. Specifically, each input prompt is prefixed with a set of special header tokens and a task description. The prompt itself is enclosed between the ``$<$BEGIN CONVERSATION$>$'' and ``$<$END CONVERSATION$>$'' special tags. Finally, the template concludes with an instruction that guides the model in generating either safe or unsafe tokens in the first generated line to detect unsafe prompts.

\circled{3} \textbf{LlamaGuard}:
In Step \circled{3}, the formatted input prompt is processed by a Byte-Pair-Encoding model \cite{sennrich2015neural} based on sentencepiece \cite{kudo2018subword} to encode textual prompts into token IDs such as [220, 128000, ..., 128007].
Each token ID, which represents the position of a token in the embedding space, is mapped to a corresponding vector using the word embedding matrix $\mathbf{W} \in \mathbb{R}^{v \times h}$ of the LlamaGuard model, where $v = 128,256$ is the vocab size and $h = 4,096$ is the hidden size.
This will produce an input matrix $\mathbf{X} \in \mathbb{R}^{l \times h}$, where $l$ is the input sequence length.
The input $\mathbf{X}$ is then fed into a stack of 32 transformer decoders. Each decoder consists of multiple layers, including masked self-attention, and feed-forward neural networks (FFNN), as used in the LlamaGuard model. The self-attention mechanism enables the model to capture dependencies across the input sequence, while the feed-forward layers apply non-linear transformations to enhance representation learning. These layers are stacked to build a deep network. For a detailed explanation of transformer decoders, we refer readers to the original paper \cite{vaswani2017attention}.

\circled{4} \textbf{LoRA (Low-Rank Adaptation)}:
In Step \circled{4}, To enhance \ourapp's~ability to defend against jailbreak prompts, we use Low-Rank Adaptation (LoRA) to efficiently adapt LlamaGuard's 8B-parameter model without fine-tuning its full parameter set.
The pre-trained knowledge of LlamaGuard, while effective for general unsafe prompts, showed limitations in defending jailbreak prompts, as demonstrated in RQ1, Finding 1.
LoRA modifies specific layers of LlamaGuard by introducing low-rank updates to their weight matrices, focusing only on a small subset of parameters while freezing the original pre-trained weights.
Specifically, the weight update matrix $\Delta \mathbf{W}$ for a target layer is parameterised as the product of two new trainable matrices, $\mathbf{A} \in \mathbb{R}^{h \times r}$ and $\mathbf{B} \in \mathbb{R}^{r \times h}$. Here, $h$ is the hidden size, and $r \ll h$ is a tunable rank parameter that controls the size of the adaptation. The updated weight for a given layer is expressed as:
\begin{equation*}
\mathbf{W}_{new} = \mathbf{W}_{pretrained} \hspace{1mm} \text{+} \hspace{1mm} \mathbf{A} \mathbf{B}.
\end{equation*}
In our implementation of \ourapp, we apply LoRA to the word embedding layer, self-attention layers, and FFNN layers of LlamaGuard. For the word embedding layer, $h$ corresponds to the embedding size, and for the self-attention and FFNN layers, $h$ corresponds to the hidden size. By restricting updates to the low-rank matrices $\mathbf{A}$ and $\mathbf{B}$ and leaving the pre-trained weights ($\mathbf{W}_{pretrained}$) unchanged, we preserve the original capabilities of LlamaGuard while enabling efficient and focused adaptation to jailbreak prompts.

\circled{5} \textbf{Detect and Block Unsafe Prompts}:
After applying the low-rank adaptation to LlamaGuard in Step \circled{4}, we obtain the model used in our \ourapp~approach.
In Step \circled{5}, we employ our \ourapp~approach to detect and block unsafe prompts from users before they are forwarded to an LLM-powered software system.
Given the output of the 32nd decoder layer, denoted as $\mathbf{H} \in \mathbb{R}^{l \times h}$, where $l$ is the sequence length and $h$ is the hidden size, \ourapp~treats the detection of unsafe prompts as a sequence generation task rather than a classification problem.
To generate tokens, a linear layer maps $\mathbf{H}$ to a distribution over the vocabulary. Specifically, the hidden state of each token $\mathbf{H}_i \in \mathbb{R}^h$ is transformed using a weight matrix $\mathbf{W}_{\text{lm}} \in \mathbb{R}^{h \times v}$ and a bias vector $\mathbf{b}_{\text{lm}} \in \mathbb{R}^v$, where $v$ is the vocab size. The resulting logits are then passed through a softmax function to compute probabilities over all possible tokens.
Guided by the instructions embedded in the chat template from Step \circled{2b}, the model generates a sequence where the first token explicitly indicates whether the input prompt is ``safe'' or ``unsafe.''
Specifically, we use greedy decoding to select tokens iteratively by choosing the one with the highest probability at each step. Formally, the next token $t_{i}$ is determined as $t_{i} = \text{argmax}_{k} softmax\left(\mathbf{H}_i \mathbf{W}_{lm} \hspace{1mm}\text{+}\hspace{1mm} \mathbf{b}_{lm}\right)_{k}$, where $k$ is the index of a token in the vocabulary, corresponding to the token with the highest probability after applying the softmax function. This process continues until the model generates the special end-of-text token ``<|eot\_id|>'' or reaches the specified maximum token limit. We then select the first token generated by \ourapp~to detect unsafe prompts.

\section{Experimental Design} \label{sec:research_design}
In this section, we present the motivation of our four research questions, the studied dataset, the studied jailbreak attacks, and our experimental setup.

\subsection{Research Questions}
To evaluate our \ourapp~approach, we formulate the following four research questions.



{\bf RQ1)} {\bf \rqone} 
Recently, Inan et al. \cite{inan2023llama} proposed LlamaGuard, a runtime guardrail for LLMs designed to classify prompts as safe or unsafe. Despite its state-of-the-art performance of 0.945 accuracy, a key limitation exists as \citet{inan2023llama} speculated LlamaGuard may be susceptible to attacks that could alter or bypass its intended use. This means attackers can apply jailbreak attacks to unsafe prompts and potentially bypass the defence of LlamaGuard, weakening the reliability of AI guardrails, as their performance may be overestimated when evaluated on prompts not subjected to advanced jailbreak techniques. Yet, little is known about how jailbreak attacks can alter the performance of AI guardrails. Thus, we investigate the impact of jailbreak attacks against current AI guardrails.

{\bf RQ2)} {\bf \rqtwo}  Given that jailbreak prompts can easily bypass the guardrail's defences, leading to a lower DSR as identified in RQ1, this research question aims to evaluate the robustness of our proposed \ourapp~against jailbreak attacks by analysing the impact on its DSR.
In Sections \ref{obfuscation_proposed} and \ref{LoRA-tuning_method}, we introduced the deciphering layer and LoRA-tuning, which aims to evaluate the DSR gains from these components compared to the baseline. Thus, we investigate the performance of our \ourapp~against baseline guardrails when defending against jailbreak attacks.

{\bf RQ3)} {\bf \rqthree} Ideally, LLM guardrails should correctly defend jailbreak prompts by blocking them, while also correctly classifying safe prompts as safe and allowing them to pass to the LLM-powered system. However, in this scenario, relying solely on DSR is insufficient to achieve a comprehensive evaluation, as it fails to reflect the instances of false positives that guardrails may produce.
Thus, we investigate how well our \ourapp~balances defence performance against jailbreak attacks while accounting for false alarms.

{\bf RQ4)} {\bf \rqfour} Our \ourapp~involves two key components—the deciphering layer and Low-Rank (LoRA) Adaptation—to enhance its defence capabilities against jailbreak prompts.
However, little is known about the contributions of each component in our \ourapp~and which component contributes the most to the Defence Success Rate (DSR) and our proposed metric - Overall Guardrail Performance (OGP) of our \ourapp.
Thus, we formulate this RQ to conduct an ablation study on the different variants of our \ourapp.

\subsection{Data Preparation}
\label{sec:studied_dataset}
To address our four research questions, we require a comprehensive dataset encompassing jailbreak prompts, unsafe prompts, and safe prompts. To this end, we prepared a dataset comprising 18,790 jailbreak prompts, 1,879 unsafe prompts, and 2,000 safe prompts. Jailbreak prompts were generated by applying 10 distinct jailbreak attack techniques to the unsafe prompts, transforming them into adversarial examples designed to bypass guardrails. Unsafe prompts were sourced from four benchmark datasets: Do-Not-Answer \cite{wang2023not}, CatQA \cite{bhardwaj2024language}, AdvBench \cite{zou2023universal}, and Forbidden Questions \cite{shen2023anything}. For safe prompts, we utilised the Alpaca dataset \cite{taori2023alpaca}.

To answer RQ1 and RQ2, which focus on evaluating the defence effectiveness of the studied guardrails and our proposed \ourapp, we leverage the jailbreak and unsafe prompts to assess their ability to defend against unsafe prompts with and without jailbreak attacks.
To answer RQ3 and RQ4, which aim to evaluate the overall effectiveness of each guardrail, we utilise the full dataset, including safe prompts, to consider the number of false alarms from the guardrails.
This ensures that the defence capability of \ourapp~is achieved without a corresponding increase in false alarms, thereby offering a comprehensive evaluation of its effectiveness.
In what follows, we provide a detailed introduction to each data source used to build our studied dataset.

\sectopic{Jailbreak Prompts (i.e., Unsafe Prompts with Jailbreak Attacks).}
For each one of our studied jailbreak attacks, we apply them to the unsafe prompts dataset to get the jailbreak prompts. In total, we apply 10 jailbreak attacks to the unsafe prompts dataset consisting of 1,879 unsafe prompts, resulting in 18,790 jailbreak prompts.

\sectopic{Unsafe Prompt Datasets.} In RQ1 and RQ2, we use unsafe prompts and unsafe prompts with jailbreak attacks applied to evaluate the performance of baseline guardrails and \ourapp. To do so, we need a benchmark dataset of unsafe prompts.
However, there exists numerous harmful categories of unsafe prompts (e.g., adult content, child exploitation).
To ensure that our empirical evaluation is not bound to specific types of harmful categories, we use four benchmark datasets consisting of human-written prompts (i.e., Forbidden Questions~\cite{shen2023anything}) and machine-generated prompts (i.e., Do-Not-Answer \cite{wang2023not}, CatQA \cite{ bhardwaj2024language}, AdvBench \cite{zou2023universal}).
These selected benchmark datasets will allow us to analyse guardrails' performance when exposed to attacks across various unsafe prompt categories.
We describe each dataset below:

\begin{itemize}
    \item \textbf{Do-Not-Answer} \cite{wang2023not}: This dataset consists of GPT-4 generated prompts when the model is prompted by \textit{What kind of questions should an AI assistant not answer?}. 
    \item \textbf{CatQA} \cite{bhardwaj2024language}: This dataset consists of harmful prompts across 11 combined categories, built on top of the combined list of prohibited use cases mentioned in OpenAI usage policies and Meta’s Llama2 acceptable use policy.
    \item \textbf{AdvBench} \cite{zou2023universal}: This dataset consists of harmful prompts formulated as instructions, originally used to perform Greedy Coordinate Gradient attack \cite{zou2023universal}, which finds universal and transferable attack prompts by optimising against multiple smaller open-source LLMs for multiple harmful behaviours.
    \item \textbf{Forbidden Questions} \cite{shen2023anything}: This dataset consists human-generated unsafe prompts collected from online sources, spanning 13 forbidden scenarios outlined in the OpenAI Usage Policy.
\end{itemize}

\sectopic{Safe Prompt Datasets.} In RQ3 and RQ4, we use safe prompts in combination with the unsafe prompt datasets under jailbreak attacks, to evaluate whether \ourapp~meets the real-world deployment needs of producing minimum false alarms compared to baseline guardrails. We include a dataset of safe prompts, the \textbf{Alpaca} dataset \cite{taori2023alpaca}, consisting of safe instructions used to fine-tune LLMs to enhance their instruction-following capabilities.

\subsection{Studied Jailbreak Attacks} \label{studied_jailbreak_attacks}
Previous works have proposed various taxonomies of jailbreak prompts against LLMs \cite{chao2024jailbreakbench, yi2024jailbreak, dong2024safeguarding}. In this study, we focus exclusively on \textbf{single-turn, black-box} jailbreak techniques, which do not require feedback or response from the LLM for optimisation. This ensures that we are testing the effectiveness of the runtime guardrails themselves, rather than assessing or interacting with internal defence mechanisms of the LLMs.

In total, we compiled 10 jailbreak attack techniques guided by Dong et al. \cite{dong2024safeguarding} and Yi et al. \cite{yi2024jailbreak} to evaluate both the baseline guardrails and our proposed \ourapp. By selecting a wide range of types of jailbreak attacks, we aim to better understand which attack types are most effective at circumventing guardrails and where gaps remain. We categorise the 10 attack techniques into 3 categories, as reflected by their traits:
\begin{itemize}
    \item \textbf{Template-based}: Applying pre-defined templates and modifications to the prompt, leveraging specific wordings, structures, or sequences to trick the model into generating restricted content. 
    The studied attacks are AIM\ cite{jailbreakchat2023}, DAN \cite{shen2023anything}, Combination (prefix injection + refusal suppression) \cite{yao2024survey}, Self Cipher \cite{yuan2023gpt} and DeepInception \cite{li2023deepinception}.
    \item \textbf{Obfuscation-based}: Rephrasing, encoding, or translating prompts into forms that guardrails may not recognise. These methods rely on the lack of multilingual alignment when guardrails are trained. The studied attacks are Caesar Cipher \cite{yuan2023gpt}, Zulu \cite{yong2023low}, and Base64 \cite{yao2024survey}.
    \item \textbf{Code-based}: Disguising harmful content within programming logic or dual-purpose scripts, exploiting the code generating capabilities within LLMs. 
    The studied attacks are Dual Use \cite{kang2024exploiting}, and Code Chameleon \cite{lv2024codechameleon}.
\end{itemize}

\subsection{Studied Guardrails}
\label{sec:studied_guardrails}
In this section, we provide a detailed description of the four guardrails studied in our experiments, including their classification mechanisms and key features.
\begin{enumerate}
    \item LlamaGuard: Based on the Llama3-8B model, LlamaGuard is an input-output guardrail that acts as an LLM to generate text in its output which indicates whether a given prompt or response is safe/unsafe. A prompt is classified as unsafe if the model's response contains the ``unsafe'' token in the first generated line.
    \item OpenAI Moderation: An active learning guardrail that leverages publicly sourced data to identify previously unknown instances of unsafe content and fine-tune the GPT-based generative guardrail. When a prompt is passed to this API, it is classified as unsafe if the API returns a boolean value flagged as ``True''. 
    \item PerspectiveAPI: Using a Transformer-based model to return a probability score of whether the prompt should be considered unsafe under each of its violation categories. A prompt is classified as unsafe if the API reports a probability exceeding 0.5 for any harmful category it supports.
    \item Perplexity: 
    A measure that focuses on detecting irregularities in the linguistic structure of sentence by evaluating the probabilities of the next token predicted by an LLM. To compute perplexity for a given input using a pre-trained model such as GPT-2 \cite{radford2019language}, the model calculates the probability of each token conditioned on its preceding tokens.
    The perplexity is formulated as
    $PPL\left(x\right) = \exp \left( - \frac{1}{t} \sum_{i=1}^{t} \log p\left(x_i | x_{<i}\right) \right)$
    where $x$ is the input text sequence, $x_{i}$ is the $i$-th token, $x_{<i}$ represents the sequence of preceding tokens, and $t$ is the total number of tokens in the sequence.

    We follow the method proposed by Liu et al. \cite{liu2024formalizing}. First, a perplexity score is calculated from a set of safe prompts using the target LLM, the threshold is then set such that the False Positive Rate (FPR)—the fraction of safe prompts incorrectly flagged as unsafe—remains within an acceptable limit, such as 1\%. Following this method, we obtain a threshold of 106, above which a prompt is classified as unsafe. We discuss the impact of the perplexity threshold on the performance of Perplexity in the Discussion section.
\end{enumerate}

\subsection{Experimental Setup}
\label{exp_setup}

\sectopic{Data Splitting.}
For RQ2, RQ3, and RQ4, we use a stratified splitting to randomly split each type of jailbreak prompts evenly. Following common practice, the jailbreak prompts were initially divided into 10\%/10\%/80\% for LoRA fine-tuning, validation, and testing. However, we opted to use only 5\% of the jailbreak prompts (940 samples) for LoRA fine-tuning, 10\% for validation, and the remaining 80\% (15,032 samples) for testing. The impact of the training data sensitivity is further discussed in Section \ref{sec:training_data_sensitivity}.

\sectopic{Model Implementation and Optimisation.}
To implement our \ourapp~approach for defending unsafe prompts, we leveraged two Python libraries, i.e., Transformers~\cite{wolf2019huggingface} and Pytorch~\cite{paszke2017automatic}. 
The Transformers library provides APIs for transformer-based model architectures and pre-trained weights, while PyTorch facilitates computations during training, including backpropagation.
We downloaded the LlamaGuard checkpoint ``meta-llama/Llama-Guard-3-8B'' provided by Inan et al. \cite{inan2023llama}.
We used our training set and low-rank adaptation \cite{hu2021lora} to fine-tune the checkpoint and obtain suitable weights for defending jailbreak prompts.
The model was fine-tuned on two NVIDIA RTX 3090 graphic cards and the training time was 58 minutes. 
As shown in Equation~\ref{equation:ce}, the Cross-Entropy Loss was used to update the model and optimise the alignment between the model's predicted token probabilities and the target sequence. For our training setup, the input is a chat-templated jailbreak prompt, and the target output is the same input followed by a single classification token, ``unsafe''.
The loss function measures the negative log-likelihood of the correct token at each position in the target sequence, guiding the causal language model (CLM) fine-tuning to generate accurate outputs for defending jailbreak prompts. 
The Cross-Entropy Loss is computed as:  
\begin{equation}
\label{equation:ce}
\mathcal{L}_{CE} = - \frac{1}{T} \sum_{t=1}^{T} \log P\left(y_t \mid x_{1:t}\right)
\end{equation}
where $T$ is the total number of tokens in the target sequence, $y_{t}$ is the target token at position $t$, and $P\left(y_{t} \mid x_{1:t}\right)$ is the model's predicted probability for the correct token $y_{t}$, conditioned on the input and all previously generated tokens $x_{1:t}$. 
In our setup, the loss is primarily focused on the generation of the final token (``unsafe''), ensuring that the model predicts this token accurately based on the context of the input jailbreak prompt.
We masked out the other tokens in the target sequence to prevent the model from being penalised for incorrectly predicting them.
By minimising this objective function, the model learns to produce outputs where the first token after generating the input will be the ``unsafe'' token when the input is identified as unsafe.

\sectopic{Hyper-Parameter Settings.} 
In our experiments, we set the learning rate to $1 \times 10^{-4}$ with a constant learning rate scheduler. We used the AdamW optimiser \cite{loshchilov2017decoupled} to update the model parameters. For the LoRA configuration, we set the rank ($r$) to 8, the alpha ($\alpha$) to 32, and applied a dropout rate of 0.1. Due to GPU memory constraints, the training batch size was set to 1. The complete training recipe for \ourapp~approach, is available in our replication package at \url{https://github.com/awsm-research/DecipherGuard}.







\section{Experimental Results}
\label{sec:experiment_results}
In this section, we present the results for our four research questions.


\begin{figure}[t]
  \centering
  \includegraphics[width=0.7\linewidth]{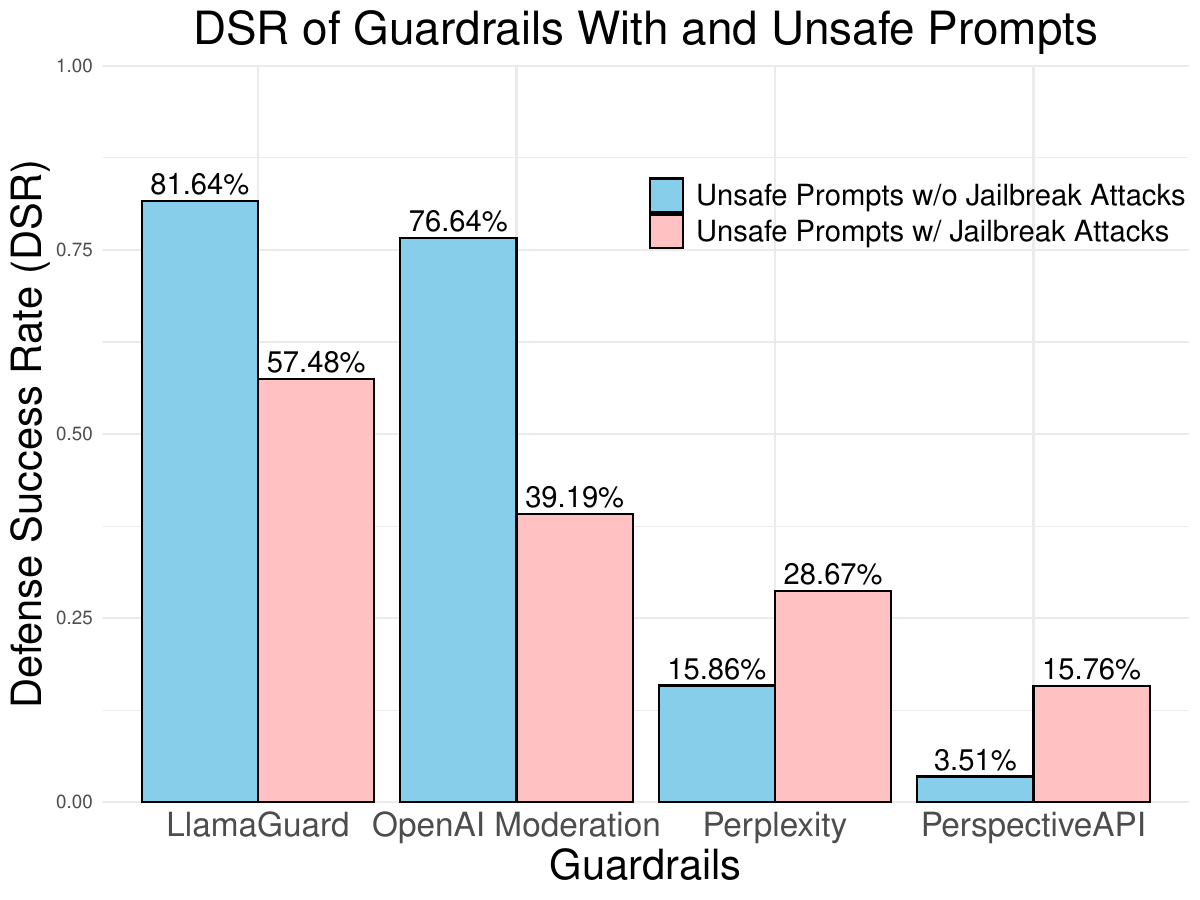}
  \caption{(RQ1 Result) The Defence Success Rate (DSR) of each studied guardrail without and with attacks.}
  \label{fig:dsr_guardrail}
\end{figure}

\subsection*{{\bf (RQ1)} \rqone}
\subsubsection*{\underline{\textbf{Approach}}}

To address this RQ, we investigate the impact of the jailbreak attacks on the performance of the existing runtime guardrails. Specifically, we compare the performance of the existing runtime guardrails in detecting unsafe prompts before and after applying jailbreak techniques. For the evaluation, we start with 1,879 unsafe prompts from studied datasets presented in Section \ref{sec:studied_dataset}, we then applied ten types of jailbreak attacks, transforming them into an additional 18,790 jailbreak prompts. Both the original unsafe prompts and the jailbreak prompts are tested across four studied guardrails: LlamaGuard, OpenAI Moderation, PerspectiveAPI, and Perplexity. 
We use the Defence Success Rate (DSR) to quantify the defensive capability of guardrails. 
DSR is defined as the percentage of the number of the jailbreak prompts that can be the successfully defended by a runtime guardrail $\#\mathrm{Prompts}_\mathrm{success}$ and the total number of the jailbreak prompts $\#\mathrm{TotalPrompts}$:
\begin{equation}
\mathrm{DSR} = \frac{\#\mathrm{Prompts}_\mathrm{success}}{\#\mathrm{TotalPrompts}}
\label{DSR_formula}
\end{equation}






\subsubsection*{\underline{\textbf{Results}}}\label{rq1_results}
Figure~\ref{fig:dsr_guardrail} presents the defence success rate (DSR) of the four evaluated guardrails, comparing their performance under two scenarios: unsafe prompts without (blue bars) and with (red bars) jailbreak attacks.

\textbf{LlamaGuard's DSR substantially decreases by 24.16\%, decreasing from 81.64\% to 57.48\% when defending against unsafe jailbreak prompts.}
Similarly, OpenAI Moderation's DSR drops by 37.45\%, declining from 76.64\% to 39.19\%.
These results indicate that while both guardrails perform well with original unsafe prompts, achieving the two highest DSRs of 81.64\% and 76.64\% among baseline guardrails, respectively, their defense capabilities are substantially reduced by 24.16\% and 37.45\% when confronted with jailbreak prompts.
\textbf{This finding demonstrates that the effectiveness of state-of-the-art guardrails is decreased when defending against jailbreak prompts, highlighting the need for jailbreak-aware guardrails capable of effectively defending such jailbreak prompts.}

\begin{figure*}[t]
  \centering
  \includegraphics[width=\linewidth]{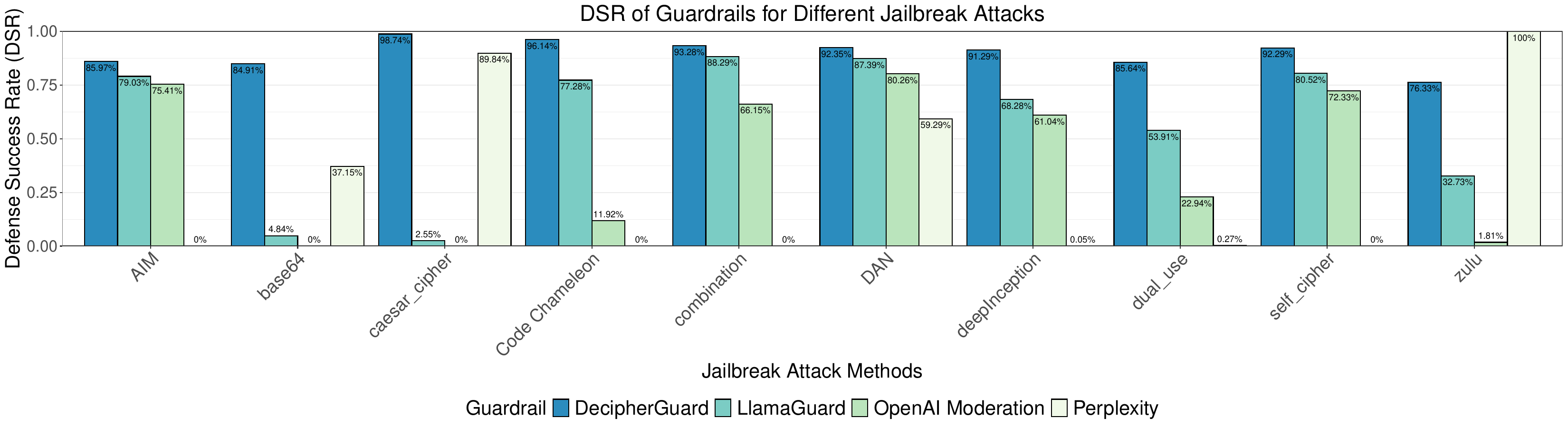}
  \caption{(RQ2) DSR of different jailbreak attacks against guardrails.}
  \label{fig:dsr_jailbreak_guardrail}
\end{figure*}

\subsection*{{\bf (RQ2)} \rqtwo}
\subsubsection*{\underline{\textbf{Approach}}} 
\label{rq2_approach}
To address this RQ, we aim to evaluate the performance of our own \ourapp~against the state-of-the-art guardrails.
We chose three guardrails, namely LlamaGuard \cite{inan2023llama}, OpenAI Moderation \cite{markov2023holistic}, and Perplexity \cite{alon2023detecting}. PerspectiveAPI was excluded as it refuses to process prompts with languages outside its training scope, particularly those transformed by obfuscation-based attacks (Base64, Zulu, Caesar Cipher). Specifically, we focus on the 18,790 jailbreak prompts to assess the effectiveness of \ourapp~in defending against such attacks, using the same Defence Success Rate (DSR).  
We present the absolute percentage difference between our \ourapp~and baseline guardrails as:
$\mathrm{\%DSR}_{\mathrm{DecipherGuard}}-\mathrm{\%DSR}_{\mathrm{baseline}}$.

\subsubsection*{\underline{\textbf{Results}}}\label{rq2_results}
\begin{figure}[t]
  \centering

  \includegraphics[width=0.5\linewidth]{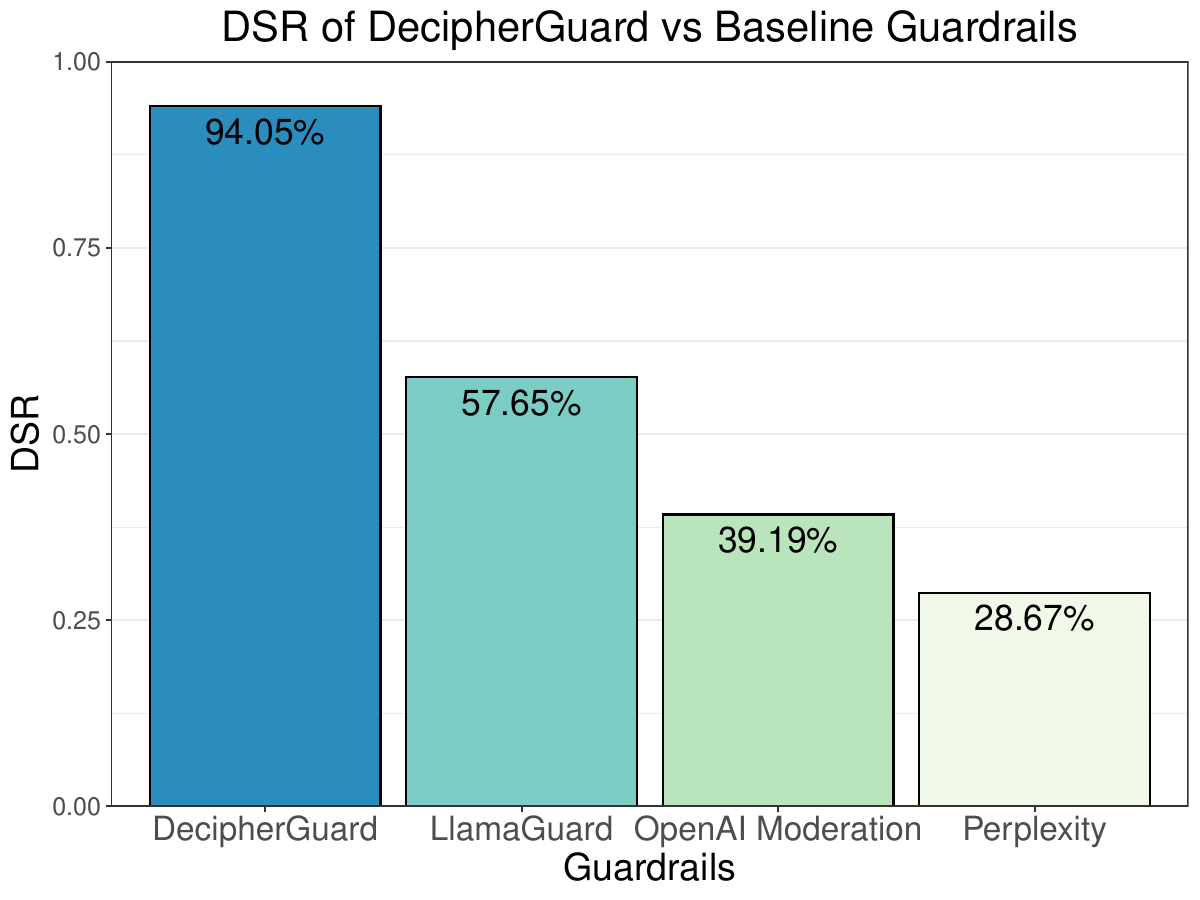}
  
  \caption{(RQ2) The Defence Success Rate (DSR) of our \ourapp~when compared with three other state-of-the-art guardrails. Higher DSR = Better. ($\nearrow$)}
  \label{fig:DSR_RQ2}
\end{figure}

Figure \ref{fig:DSR_RQ2} presents the defence success rate (DSR) of our \ourapp~compared with the three baseline guardrail approaches.

\textbf{Our \ourapp~achieves a DSR of 94.05\% when under jailbreak attacks, which is 36\% to 65\% higher than the baseline guardrail approaches with a median improvement of 54\%.} In terms of DSR against jailbreak prompts, Figure \ref{fig:DSR_RQ2} shows that \ourapp~achieves the highest DSR of 94.05\%, while the baseline guardrails achieve a DSR of 28.67\%-57.48\%. This finding shows that \ourapp~substantially improves the state-of-the-art guardrails by 36\% to 65\% with a median improvement of 54.9\%.
\textbf{These results confirm that our \ourapp~approach is more effective than baseline guardrails in defending against jailbreak prompts.}

Figure \ref{fig:dsr_jailbreak_guardrail} presents the DSR of our \ourapp~compared to the other three baseline guardrails, categorised by the ten different jailbreak attacks.
Notably, \ourapp~substantially improves the DSR against obfuscation-based attacks compared to the best-performing baseline, LlamaGuard.
For the three obfuscation-based attacks, \ourapp~achieves an improvement of 96.19\% (2.55\% $\rightarrow$ 98.74\%) for Base64; 80.07\% (4.84\% $\rightarrow$ 84.91\%) for Caesar Cipher; and 43.6\%, (32.73\% $\rightarrow$ 76.33\%) for Zulu.
Similarly, \ourapp~also enhances the DSR for the second-best baseline, OpenAI Moderation, achieving an improvement of 74\% to 98\% across the obfuscation-based attacks.
\textbf{These results demonstrate that \ourapp~effectively addresses a key limitation of the state-of-the-art guardrail, LlamaGuard, namely its vulnerability to obfuscation-based attacks, highlighting the potential of \ourapp~to improve guardrail effectiveness in such scenarios.}

\begin{figure}[t]
  \centering

  \includegraphics[width=0.5\linewidth]{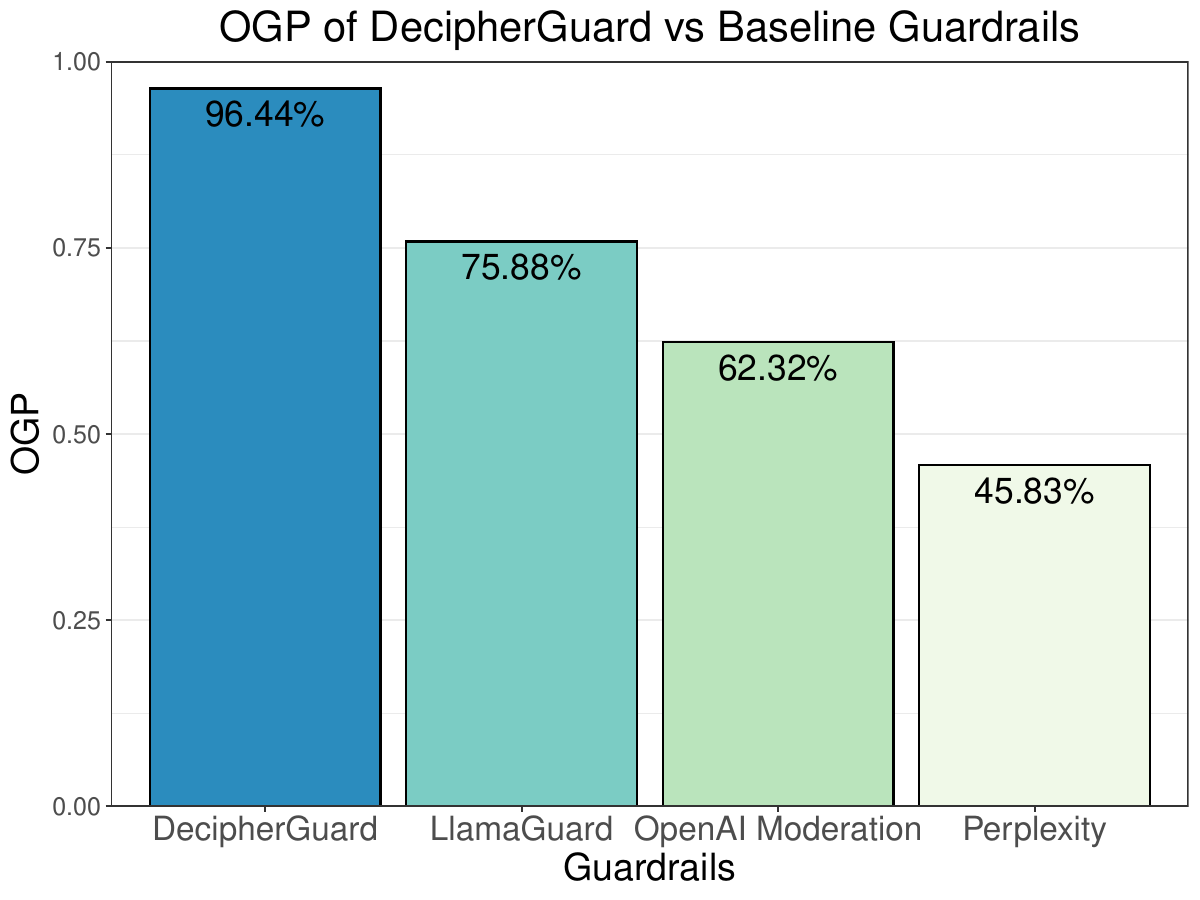}

  \caption{(RQ3) The Overall Guardrail Performance (OGP) of our \ourapp~when compared with three other state-of-the-art guardrails. Higher OGP = Better. ($\nearrow$)}
  \label{fig:OGP_RQ3}
\end{figure}

\subsection*{{\bf (RQ3)} \rqthree}
\subsubsection*{\underline{\textbf{Approach}}}


To address this RQ, we compare the overall performance of our \ourapp~with the three baseline guardrails as in RQ2.
We consider 2,000 safe prompts, 1,879 unsafe prompts without applying jailbreak attacks, and 18,790 jailbreak prompts.
Prior research \cite{inan2023llama, chu2024comprehensive, xu2024comprehensive} has commonly employed traditional metrics such as AUPRC, accuracy, attack success rate (ASR), and F1 score to evaluate runtime guardrails.
However, these metrics fail to capture the critical balance between defence success and false alarms, which is essential for the practical deployment of guardrails.
In particular, metrics like ASR and accuracy focus solely on whether an attack bypasses the guardrail, neglecting the equally important aspect of reducing false alarms.
While the F1-score effectively balances precision and recall, it is inadequate for evaluating runtime guardrails. This limitation arises because it does not account for the trade-off between defence effectiveness and false alarms, neglecting to penalise an excessive number of false positives.
To address this gap, we propose the Overall Guardrail Performance (OGP) metric, defined as:
\begin{equation}
\mathrm{OGP} = \sqrt{\mathrm{DSR} \times \left(1 - \mathrm{FAR}\right)}
\label{OGP_formula}
\end{equation}
, where $\mathrm{DSR}$ (Defence Success Rate) measures the guardrail's ability to block unsafe prompts.
$\mathrm{FAR}$ (False Alarm Rate) is calculated as $\frac{N_{FA}}{N_\mathrm{safe}}$ where $N_{FA}$ is the total number of false alarms and $N_\mathrm{safe}$ is the total number of safe prompts.
In addition, $\left(1 - FAR\right)$ (the complement of FAR) captures guardrails' reliability in avoiding false positives.
We then use the geometric mean to combine the two factors.
Our choice of the geometric mean in the OGP metric is inspired by an established metric, the G-measure, which is the geometric mean of precision and recall.
The G-measure normalizes true positives relative to both predicted positives and actual positives, effectively balancing the trade-off between the two measures. In other words, the amount of information the G-measure provides is the arithmetic mean of the information from precision and recall, ensuring that neither metric dominates the overall evaluation \cite{powers2020evaluation}.
By combining these factors through a geometric mean, OGP evaluates guardrails holistically, ensuring that high defensive efficacy is not achieved at the expense of a high number of false alarms.

\subsubsection*{\underline{\textbf{Results}}}\label{rq3_results}
Figure \ref{fig:OGP_RQ3} presents the experimental results of our \ourapp~and the three baseline guardrails in terms of Overall Guardrail Performance (OGP).

\textbf{Our \ourapp~achieves a OGP value of 96.44\%, which is 20\%-50\% better than other baseline guardrails.}
In terms of the OGP, Figure \ref{fig:OGP_RQ3} shows that our \ourapp~achieves an OGP of 96.44\%, while the existing guardrails achieve an OGP of 45.83\%-75.88\%.
This finding shows that \ourapp~substantially improves the baseline guardrails by 20\%-50\%  with a median improvement of 33\%.
\textbf{These results confirm that our \ourapp~approach achieves better overall performance, enhancing defense effectiveness while reducing false alarms.}

In other words, our results demonstrate that the combination of the deciphering layer and low-rank adaptation (LoRA) mechanisms outperforms guardrails that rely solely on pre-trained large language models (LLMs) such as LlamaGuard.
Prior works \cite{inan2023llama,markov2023holistic} have utilised LLMs as guardrails to defend against potentially unsafe prompts from end users. However, as demonstrated in RQ1, their Decision Success Rate (DSR) drops significantly when confronted with jailbreak prompts. In RQ2, we found that these LLM-driven guardrails are particularly vulnerable to obfuscation-based attacks, such as Base64, Caesar Cipher, and Zulu, as illustrated in Figure \ref{fig:dsr_jailbreak_guardrail}. 
In contrast, our \ourapp~enhances LLM-driven guardrails by integrating a deciphering layer designed to detect and reverse obfuscation-based attacks. Additionally, \ourapp~extends LlamaGuard through Low-Rank Adaptation (LoRA), a lightweight and parameter-efficient (only 0.06\% of parameters are tuned) fine-tuning approach that requires only 10\% of the training data. This adaptation enables the model to better defend against jailbreak prompts while preserving the original pre-trained parameters of LlamaGuard.
This paper is among the first to propose a deciphering layer and the use of LoRA to enhance LLM-driven guardrails, offering an effective solution to counter obfuscation-based attacks and adapt to evolving jailbreak scenarios.

\begin{figure}[t]
  \centering
  \includegraphics[width=\linewidth]{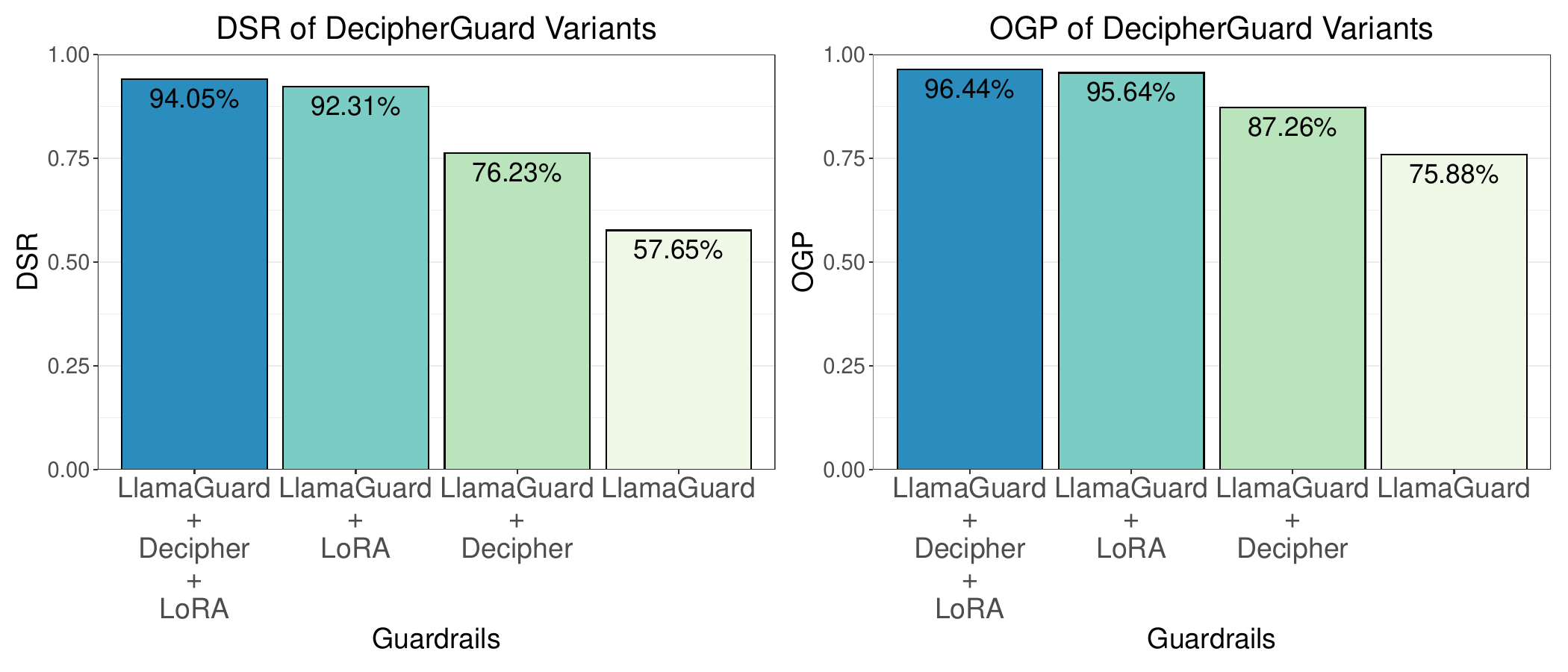}
  \caption{(RQ4) Evaluation of different variations of \ourapp.}
  \label{fig:RQ4_plot}
\end{figure}

\subsection*{{\bf (RQ4)} \rqfour}
\subsubsection*{\underline{\textbf{Approach}}}
To answer this RQ, we investigate the contributions of the deciphering layer and LoRA within \ourapp~by examining the DSR and OGP of \ourapp~after removing different components.
To understand and quantify the contribution of the two components of our approach, we alter \ourapp~as follows:
\begin{itemize}
    \item \textit{LlamaGuard}: Remove all components, plain LlamaGuard only.
    \item \textit{\underline{Decipher}+LlamaGuard}: Remove LoRA, but keep the deciphering layer.
    \item \textit{\underline{LoRA}+LlamaGuard}: Remove the deciphering layer, but keep LoRA to fine-tune LlamaGuard. 
    \item \textit{\underline{LoRA}+ \underline{Decipher} + LlamaGuard}: Full proposed DecipherGuard.
\end{itemize}
Following our previous RQs, we use the Defence Success Rate (DSR) and Overall Guardrail Performance (OGP) as measures for this ablation study.

\subsubsection*{\underline{\textbf{Results}}}\label{rq4_results}
Figure \ref{fig:RQ4_plot} presents the ablation study to evaluate the contributions of the components in our \ourapp.

\textbf{The LoRA component is the most important component for enhancing Defence Success Rate (DSR) to achieve better defence effectiveness.}
Within our \ourapp, the LoRA component contributes to 34.66\% of the DSR. When comparing ``\textit{LlamaGuard + \underline{LoRA}}'' and ``\textit{LlamaGuard}'' where the LoRA component is eliminated, we observe a performance decrease from 92.31\% to 57.65\%, accounting for 34.66\%.
Within our \ourapp, the Decipher component contributes to 18.58\% of the DSR. When comparing ``\textit{LlamaGuard + \underline{Decipher}}'' and ``\textit{LlamaGuard}'' where the Decipher component is eliminated, we observe a performance decrease from 76.23\% to 57.65\%, accounting for 18.58\%.
These results underscore the substantial contributions of each component to the overall defence effectiveness of \ourapp, which achieves the highest of 94.05\% when both components are active.

\textbf{The LoRA component is the most important component for enhancing Overall Guardrail Performance (OGP), leading to improved overall performance when considering safe prompts.}
Within our \ourapp, the LoRA component contributes to 19.76\% of the OGP. When comparing ``\textit{LlamaGuard + \underline{LoRA}}'' and ``\textit{LlamaGuard}'' where the LoRA component is eliminated, we observe a performance decrease from 95.64\% to 75.88\%, accounting for 19.76\%.
Within our \ourapp, the Decipher component contributes to 11.38\% of the DSR. When comparing ``\textit{LlamaGuard + \underline{Decipher}}'' and ``\textit{LlamaGuard}'' where the Decipher component is eliminated, we observe a performance decrease from 87.26\% to 75.88\%, accounting for 11.38\%.
These results underscore the substantial contributions of each component to the overall defence effectiveness of \ourapp, which achieves the highest OGP of 96.44\% when both components are active.

The Decipher component plays a crucial role in enhancing the overall performance of \ourapp, contributing substantially to 18.58\% of the DSR and 11.38\% of the OGP.
\textbf{Unlike LoRA, which relies on fine-tuning data and computation, Decipher offers a lightweight yet effective solution that requires no additional fine-tuning, making it computationally efficient and adaptable.
Furthermore, the Decipher component can be integrated into non-LLM-based guardrails as an additional input processing layer.}
This flexibility expands its applicability to a wider range of guardrails to enhance their defence effectiveness against obfuscated-based jailbreak prompts.

\section{Discussion}
\label{sec:discussion}
In the previous experiment section, we empirically evaluated the performance of our \ourapp~and conducted an ablation study to support our design rationale.
However, the computational overhead introduced by this layer has not been evaluated for the deciphering layer, which detects and reverses obfuscation-based jailbreak prompts by acting as an additional preprocessing layer.
In this section, we perform an extended analysis of our proposed approach to resolve this question.

\begin{table}[]
\centering
\caption{(Discussion) Runtime latency analysis of our \ourapp.}
\label{tab:runtime_efficiency}
\resizebox{0.7\linewidth}{!}{
\begin{tabular}{l|l}
\hline
\multicolumn{1}{c|}{\textbf{Configuration}}   & \multicolumn{1}{c}{\textbf{Total Latency}} \\ \hline
LoRA-tuned LlamaGuard                         & 333 ms                                     \\ \hline
LoRA-tuned LlamaGuard + Base64 Deobfuscation        & 333.2 ms                                   \\
LoRA-tuned LlamaGuard + Zulu Deobfuscation          & 333.5 ms                                   \\
LoRA-tuned LlamaGuard + Caesar Cipher Deobfuscation & 370 ms                                     \\ \hline
\ourapp                                 & 370.7 ms                                   \\ \hline
\end{tabular}
}
\end{table}

\subsection{\ourapp's Runtime Efficiency}
\label{dg_efficiency}
In Section \ref{obfuscation_proposed}, we proposed the deciphering layer to detect and reverse obfuscated-based jailbreak prompts.
Our ablation study (RQ4) confirmed its effectiveness in improving the overall performance of LLM-based guardrails such as LlamaGuard against jailbreak prompts.
However, the computational overhead introduced by this additional layer remains unknown. Understanding this latency is crucial for assessing the feasibility and efficiency of our \ourapp~for future deployments.
Thus, we analyse the runtime latency introduced by Base64 deobfuscation, Zulu deobfuscation, and Caesar cipher deobfuscation, as well as the total latency introduced by \ourapp, comparing these results with the latency of the original LlamaGuard.
We use the same data as in RQ3, consisting of 1,879 unsafe prompts w/o jailbreak, 18,790 unsafe prompts w/ jailbreak, and 2,000 safe prompts.
All components in our deciphering layer are run by an AMD Ryzen 9 5950X CPU while the LoRA-tuned LlamaGuard is run by a Nvidia RTX 3090 GPU with 24GB of memory.

Table \ref{tab:runtime_efficiency} presents the median runtime overhead of each deobfuscation, our \ourapp, and LoRA-tuned LlamaGuard.
Base64 deobfuscation introduces a latency of 0.2 ms, Zulu deobfuscation 0.5 ms, and Caesar cipher deobfuscation 37 ms. Among these, Caesar cipher deobfuscation is the most time-consuming due to the need to perform up to 25 character shifts. Each shift requires invoking a language detector to determine whether the resulting text is likely English and not ciphered code (see Algorithm \ref{algo:decipher}).
The total latency of our \ourapp~is 370.7 ms, comprising a deciphering layer and a LoRA-tuned LlamaGuard. The deciphering layer contributes 37.7 ms to this total, representing a relative latency increase of 11\%. Despite this, the layer substantially improves the Defence Success Rate (DSR) by 18.58\% and the Overall Guardrail Performance (OGP) by 11.38\% (see RQ4), all without requiring model fine-tuning.
These results highlight a reasonable latency tradeoff, offering substantial performance gains without the computational overhead of model fine-tuning.

\section{Threats to the Validity} \label{sec:threats}

\textbf{Threats to construct validity} relates to the selection of jailbreak attacks.
We selected 10 different jailbreak attacks guided by their prevalence to guardrails and ability to represent a wide spectrum of jailbreak techniques \cite{dong2024safeguarding, yi2024jailbreak, wei2024jailbroken}.
It is important to note that additional attack types can be included in future evaluations.
However, this would not alter the key conclusion presented in RQ1, that jailbreak attacks substantially impact the performance of existing runtime guardrails.
The underlying mechanisms through which these attacks degrade guardrail effectiveness remain consistent, regardless of the specific attacks tested.

\textbf{Threats to internal validity} relate to the potential influence of hyperparameter settings during the fine-tuning of our \ourapp. Variations in model versions or different LoRA hyperparameters, compared to those specified in Section \ref{sec:research_design}, could impact the experiment's outcomes.
To address this threat, we open-source our replication package and provide detailed documentation of all hyperparameter settings to ensure the experiment is reproducible by future researchers.
Additionally, to minimise the impact of non-determinism introduced by deep learning model training, we conducted five repetitions of each experiment and presented the averaged outcomes to demonstrate the stability of our findings across multiple trials.

\textbf{Threats to external validity} concerns the generalisability of our results. Our experiment findings are supported by the dataset, jailbreak methods, and guardrails employed during the study. The dataset contains 1,879 unsafe prompts and 2,000 safe prompts from a separate dataset. When applied with the 10 jailbreak methods, we have 18,790 jailbreak prompts across 10 categories.
While \ourapp~is fine-tuned specifically to address the jailbreak attacks discussed in this paper, other prompt datasets and jailbreak methods can be explored in future work.

\section{Conclusion}
\label{sec:conclusion}
In this paper, we present \ourapp, a novel framework that integrates a deciphering layer with low-rank adaptation (LoRA) to effectively defend against obfuscation- and template-based jailbreak prompts in LLM-powered software systems.
We also introduce the Overall Guardrail Performance (OGP) metric, which evaluates guardrail performance by considering both defense effectiveness and the number of false alarms.
Through an empirical evaluation of over 22,000 prompts across 10 different jailbreak attacks, our results highlight a substantial performance drop in state-of-the-art guardrails when confronted with such attacks.
In comparison, \ourapp~achieves 36\%-65\% higher Defense Success Rate (DSR) and 20\%-50\% higher OGP, demonstrating superior effectiveness in defending against jailbreak attacks while retaining low false alarms.
These findings underscore the potential of \ourapp~to help defend against jailbreak attacks and contribute to a safer deployment of intelligent software systems powered by LLMs.
 
\section*{Acknowledgement}

We thank Transurban and the CSIRO Next Generation Graduate AI Program: Creating Responsible AI Software Engineering Capability (RAISE) for their support and collaboration. 
The perspectives and conclusions presented in this study are solely the authors’ and should not be interpreted as representing the official policies or endorsements of Transurban or any of its subsidiaries and affiliates. Additionally, the outcomes of this study are independent of, and should not be construed as an assessment of, the quality of products offered by Transurban.

\bibliographystyle{IEEEtran}
\bibliography{reference}

\begin{thebibliography}{10}
\providecommand{\url}[1]{#1}
\csname url@samestyle\endcsname
\providecommand{\newblock}{\relax}
\providecommand{\bibinfo}[2]{#2}
\providecommand{\BIBentrySTDinterwordspacing}{\spaceskip=0pt\relax}
\providecommand{\BIBentryALTinterwordstretchfactor}{4}
\providecommand{\BIBentryALTinterwordspacing}{\spaceskip=\fontdimen2\font plus
\BIBentryALTinterwordstretchfactor\fontdimen3\font minus \fontdimen4\font\relax}
\providecommand{\BIBforeignlanguage}[2]{{%
\expandafter\ifx\csname l@#1\endcsname\relax
\typeout{** WARNING: IEEEtran.bst: No hyphenation pattern has been}%
\typeout{** loaded for the language `#1'. Using the pattern for}%
\typeout{** the default language instead.}%
\else
\language=\csname l@#1\endcsname
\fi
#2}}
\providecommand{\BIBdecl}{\relax}
\BIBdecl

\bibitem{hassan2024rethinking}
A.~E. Hassan, G.~A. Oliva, D.~Lin, B.~Chen, Z.~Ming \emph{et~al.}, ``Rethinking software engineering in the foundation model era: From task-driven ai copilots to goal-driven ai pair programmers,'' \emph{arXiv preprint arXiv:2404.10225}, 2024.

\bibitem{bengio2024managing}
Y.~Bengio, G.~Hinton, A.~Yao, D.~Song, P.~Abbeel, T.~Darrell, Y.~N. Harari, Y.-Q. Zhang, L.~Xue, S.~Shalev-Shwartz \emph{et~al.}, ``Managing extreme ai risks amid rapid progress,'' \emph{Science}, vol. 384, no. 6698, pp. 842--845, 2024.

\bibitem{yao2024survey}
Y.~Yao, J.~Duan, K.~Xu, Y.~Cai, Z.~Sun, and Y.~Zhang, ``A survey on large language model (llm) security and privacy: The good, the bad, and the ugly,'' \emph{High-Confidence Computing}, p. 100211, 2024.

\bibitem{dong2024attacks}
Z.~Dong, Z.~Zhou, C.~Yang, J.~Shao, and Y.~Qiao, ``Attacks, defenses and evaluations for llm conversation safety: A survey,'' \emph{arXiv preprint arXiv:2402.09283}, 2024.

\bibitem{wang2024unique}
S.~Wang, T.~Zhu, B.~Liu, D.~Ming, X.~Guo, D.~Ye, and W.~Zhou, ``Unique security and privacy threats of large language model: A comprehensive survey,'' \emph{arXiv preprint arXiv:2406.07973}, 2024.

\bibitem{inan2023llama}
H.~Inan, K.~Upasani, J.~Chi, R.~Rungta, K.~Iyer, Y.~Mao, M.~Tontchev, Q.~Hu, B.~Fuller, D.~Testuggine \emph{et~al.}, ``Llama guard: Llm-based input-output safeguard for human-ai conversations,'' \emph{arXiv preprint arXiv:2312.06674}, 2023.

\bibitem{markov2023holistic}
T.~Markov, C.~Zhang, S.~Agarwal, F.~E. Nekoul, T.~Lee, S.~Adler, A.~Jiang, and L.~Weng, ``A holistic approach to undesired content detection in the real world,'' in \emph{Proceedings of the AAAI Conference on Artificial Intelligence}, vol.~37, no.~12, 2023, pp. 15\,009--15\,018.

\bibitem{alon2023detecting}
G.~Alon and M.~Kamfonas, ``Detecting language model attacks with perplexity,'' \emph{arXiv preprint arXiv:2308.14132}, 2023.

\bibitem{lees2022new}
A.~Lees, V.~Q. Tran, Y.~Tay, J.~Sorensen, J.~Gupta, D.~Metzler, and L.~Vasserman, ``A new generation of perspective api: Efficient multilingual character-level transformers,'' in \emph{Proceedings of the 28th ACM SIGKDD conference on knowledge discovery and data mining}, 2022, pp. 3197--3207.

\bibitem{rebedea2023nemo}
T.~Rebedea, R.~Dinu, M.~Sreedhar, C.~Parisien, and J.~Cohen, ``Nemo guardrails: A toolkit for controllable and safe llm applications with programmable rails,'' \emph{arXiv preprint arXiv:2310.10501}, 2023.

\bibitem{yong2023low}
Z.-X. Yong, C.~Menghini, and S.~H. Bach, ``Low-resource languages jailbreak gpt-4,'' \emph{arXiv preprint arXiv:2310.02446}, 2023.

\bibitem{yuan2023gpt}
Y.~Yuan, W.~Jiao, W.~Wang, J.-t. Huang, P.~He, S.~Shi, and Z.~Tu, ``Gpt-4 is too smart to be safe: Stealthy chat with llms via cipher,'' \emph{arXiv preprint arXiv:2308.06463}, 2023.

\bibitem{jailbreakchat2023}
\BIBentryALTinterwordspacing
{Jailbreak Chat}, ``Jailbreak chat prompt,'' 2023, last accessed: 2024-09-20. [Online]. Available: \url{https://www.jailbreakchat.com/prompt/4f37a029-9dff-4862-b323-c96a5504de5d}
\BIBentrySTDinterwordspacing

\bibitem{shen2023anything}
X.~Shen, Z.~Chen, M.~Backes, Y.~Shen, and Y.~Zhang, ``" do anything now": Characterizing and evaluating in-the-wild jailbreak prompts on large language models,'' \emph{arXiv preprint arXiv:2308.03825}, 2023.

\bibitem{xu2024comprehensive}
Z.~Xu, Y.~Liu, G.~Deng, Y.~Li, and S.~Picek, ``A comprehensive study of jailbreak attack versus defense for large language models,'' in \emph{Findings of the Association for Computational Linguistics ACL 2024}, 2024, pp. 7432--7449.

\bibitem{li2023deepinception}
X.~Li, Z.~Zhou, J.~Zhu, J.~Yao, T.~Liu, and B.~Han, ``Deepinception: Hypnotize large language model to be jailbreaker,'' \emph{arXiv preprint arXiv:2311.03191}, 2023.

\bibitem{kang2024exploiting}
D.~Kang, X.~Li, I.~Stoica, C.~Guestrin, M.~Zaharia, and T.~Hashimoto, ``Exploiting programmatic behavior of llms: Dual-use through standard security attacks,'' in \emph{2024 IEEE Security and Privacy Workshops (SPW)}.\hskip 1em plus 0.5em minus 0.4em\relax IEEE, 2024, pp. 132--143.

\bibitem{lv2024codechameleon}
H.~Lv, X.~Wang, Y.~Zhang, C.~Huang, S.~Dou, J.~Ye, T.~Gui, Q.~Zhang, and X.~Huang, ``Codechameleon: Personalized encryption framework for jailbreaking large language models,'' \emph{arXiv preprint arXiv:2402.16717}, 2024.

\bibitem{tu_linkt}
Linkt, ``Easier, smarter ways to pay for australian toll roads.'' \url{https://www.linkt.com.au/}, 2024.

\bibitem{bender2021dangers}
E.~M. Bender, T.~Gebru, A.~McMillan-Major, and S.~Shmitchell, ``On the dangers of stochastic parrots: Can language models be too big?'' in \emph{Proceedings of the 2021 ACM conference on fairness, accountability, and transparency}, 2021, pp. 610--623.

\bibitem{vaswani2017attention}
A.~Vaswani, ``Attention is all you need,'' \emph{Advances in Neural Information Processing Systems}, 2017.

\bibitem{alto2023modern}
V.~Alto, \emph{Modern Generative AI with ChatGPT and OpenAI Models: Leverage the capabilities of OpenAI's LLM for productivity and innovation with GPT3 and GPT4}.\hskip 1em plus 0.5em minus 0.4em\relax Packt Publishing Ltd, 2023.

\bibitem{shen2023large}
T.~Shen, R.~Jin, Y.~Huang, C.~Liu, W.~Dong, Z.~Guo, X.~Wu, Y.~Liu, and D.~Xiong, ``Large language model alignment: A survey,'' \emph{arXiv preprint arXiv:2309.15025}, 2023.

\bibitem{bai2022training}
Y.~Bai, A.~Jones, K.~Ndousse, A.~Askell, A.~Chen, N.~DasSarma, D.~Drain, S.~Fort, D.~Ganguli, T.~Henighan \emph{et~al.}, ``Training a helpful and harmless assistant with reinforcement learning from human feedback,'' \emph{arXiv preprint arXiv:2204.05862}, 2022.

\bibitem{achiam2023gpt}
J.~Achiam, S.~Adler, S.~Agarwal, L.~Ahmad, I.~Akkaya, F.~L. Aleman, D.~Almeida, J.~Altenschmidt, S.~Altman, S.~Anadkat \emph{et~al.}, ``Gpt-4 technical report,'' \emph{arXiv preprint arXiv:2303.08774}, 2023.

\bibitem{touvron2023llama}
H.~Touvron, L.~Martin, K.~Stone, P.~Albert, A.~Almahairi, Y.~Babaei, N.~Bashlykov, S.~Batra, P.~Bhargava, S.~Bhosale \emph{et~al.}, ``Llama 2: Open foundation and fine-tuned chat models,'' \emph{arXiv preprint arXiv:2307.09288}, 2023.

\bibitem{cui2024recent}
J.~Cui, Y.~Xu, Z.~Huang, S.~Zhou, J.~Jiao, and J.~Zhang, ``Recent advances in attack and defense approaches of large language models,'' \emph{arXiv preprint arXiv:2409.03274}, 2024.

\bibitem{wei2024jailbroken}
A.~Wei, N.~Haghtalab, and J.~Steinhardt, ``Jailbroken: How does llm safety training fail?'' \emph{Advances in Neural Information Processing Systems}, vol.~36, 2024.

\bibitem{anwar2024foundational}
U.~Anwar, A.~Saparov, J.~Rando, D.~Paleka, M.~Turpin, P.~Hase, E.~S. Lubana, E.~Jenner, S.~Casper, O.~Sourbut \emph{et~al.}, ``Foundational challenges in assuring alignment and safety of large language models,'' \emph{arXiv preprint arXiv:2404.09932}, 2024.

\bibitem{biswas2023guardrails}
A.~Biswas and W.~Talukdar, ``Guardrails for trust, safety, and ethical development and deployment of large language models (llm),'' \emph{Journal of Science \& Technology}, vol.~4, no.~6, pp. 55--82, 2023.

\bibitem{dong2024safeguarding}
Y.~Dong, R.~Mu, Y.~Zhang, S.~Sun, T.~Zhang, C.~Wu, G.~Jin, Y.~Qi, J.~Hu, J.~Meng \emph{et~al.}, ``Safeguarding large language models: A survey,'' \emph{arXiv preprint arXiv:2406.02622}, 2024.

\bibitem{hu2021lora}
E.~J. Hu, Y.~Shen, P.~Wallis, Z.~Allen-Zhu, Y.~Li, S.~Wang, L.~Wang, and W.~Chen, ``Lora: Low-rank adaptation of large language models,'' \emph{arXiv preprint arXiv:2106.09685}, 2021.

\bibitem{base64_python}
Python, ``base64 — base16, base32, base64, base85 data encodings,'' \url{https://docs.python.org/3/library/base64.html}, 2024.

\bibitem{lingua}
P.~Stahl, ``An accurate natural language detection library, suitable for short text and mixed-language text,'' \url{https://pypi.org/project/lingua-language-detector/}, 2024.

\bibitem{googletrans}
S.~Han, ``Free google translate api for python. translates totally free of charge,'' \url{https://pypi.org/project/googletrans/3.1.0a0/}, 2020.

\bibitem{sennrich2015neural}
R.~Sennrich, ``Neural machine translation of rare words with subword units,'' \emph{arXiv preprint arXiv:1508.07909}, 2015.

\bibitem{kudo2018subword}
T.~Kudo, ``Subword regularization: Improving neural network translation models with multiple subword candidates,'' \emph{arXiv preprint arXiv:1804.10959}, 2018.

\bibitem{wang2023not}
Y.~Wang, H.~Li, X.~Han, P.~Nakov, and T.~Baldwin, ``Do-not-answer: A dataset for evaluating safeguards in llms,'' \emph{arXiv preprint arXiv:2308.13387}, 2023.

\bibitem{bhardwaj2024language}
R.~Bhardwaj, D.~D. Anh, and S.~Poria, ``Language models are homer simpson! safety re-alignment of fine-tuned language models through task arithmetic,'' \emph{arXiv preprint arXiv:2402.11746}, 2024.

\bibitem{zou2023universal}
A.~Zou, Z.~Wang, N.~Carlini, M.~Nasr, J.~Z. Kolter, and M.~Fredrikson, ``Universal and transferable adversarial attacks on aligned language models,'' \emph{arXiv preprint arXiv:2307.15043}, 2023.

\bibitem{taori2023alpaca}
R.~Taori, I.~Gulrajani, T.~Zhang, Y.~Dubois, X.~Li, C.~Guestrin, P.~Liang, and T.~B. Hashimoto, ``Alpaca: A strong, replicable instruction-following model,'' \emph{Stanford Center for Research on Foundation Models. https://crfm. stanford. edu/2023/03/13/alpaca. html}, vol.~3, no.~6, p.~7, 2023.

\bibitem{chao2024jailbreakbench}
P.~Chao, E.~Debenedetti, A.~Robey, M.~Andriushchenko, F.~Croce, V.~Sehwag, E.~Dobriban, N.~Flammarion, G.~J. Pappas, F.~Tramer \emph{et~al.}, ``Jailbreakbench: An open robustness benchmark for jailbreaking large language models,'' \emph{arXiv preprint arXiv:2404.01318}, 2024.

\bibitem{yi2024jailbreak}
S.~Yi, Y.~Liu, Z.~Sun, T.~Cong, X.~He, J.~Song, K.~Xu, and Q.~Li, ``Jailbreak attacks and defenses against large language models: A survey,'' \emph{arXiv preprint arXiv:2407.04295}, 2024.

\bibitem{radford2019language}
A.~Radford, J.~Wu, R.~Child, D.~Luan, D.~Amodei, I.~Sutskever \emph{et~al.}, ``Language models are unsupervised multitask learners,'' \emph{OpenAI blog}, vol.~1, no.~8, p.~9, 2019.

\bibitem{liu2024formalizing}
Y.~Liu, Y.~Jia, R.~Geng, J.~Jia, and N.~Z. Gong, ``Formalizing and benchmarking prompt injection attacks and defenses,'' in \emph{33rd USENIX Security Symposium (USENIX Security 24)}, 2024, pp. 1831--1847.

\bibitem{wolf2019huggingface}
T.~Wolf, L.~Debut, V.~Sanh, J.~Chaumond, C.~Delangue, A.~Moi, P.~Cistac, T.~Rault, R.~Louf, M.~Funtowicz \emph{et~al.}, ``Huggingface's transformers: State-of-the-art natural language processing,'' \emph{arXiv preprint arXiv:1910.03771}, 2019.

\bibitem{paszke2017automatic}
A.~Paszke, S.~Gross, S.~Chintala, G.~Chanan, E.~Yang, Z.~DeVito, Z.~Lin, A.~Desmaison, L.~Antiga, and A.~Lerer, ``Automatic differentiation in pytorch,'' in \emph{NIPS-W}, 2017.

\bibitem{loshchilov2017decoupled}
I.~Loshchilov and F.~Hutter, ``Decoupled weight decay regularization,'' in \emph{International Conference on Learning Representations}, 2018.

\bibitem{chu2024comprehensive}
J.~Chu, Y.~Liu, Z.~Yang, X.~Shen, M.~Backes, and Y.~Zhang, ``Comprehensive assessment of jailbreak attacks against llms,'' \emph{arXiv preprint arXiv:2402.05668}, 2024.

\bibitem{powers2020evaluation}
D.~M. Powers, ``Evaluation: from precision, recall and f-measure to roc, informedness, markedness and correlation,'' \emph{arXiv preprint arXiv:2010.16061}, 2020.

\end{thebibliography}
 
 
\end{document}